\documentclass[12pt]{article}\usepackage[]{graphicx}\usepackage[]{color}
\makeatletter
\def\maxwidth{ %
  \ifdim\Gin@nat@width>\linewidth
    \linewidth
  \else
    \Gin@nat@width
  \fi
}
\makeatother

\definecolor{fgcolor}{rgb}{0.345, 0.345, 0.345}

\usepackage{framed}
\makeatletter
 {\par\unskip\endMakeFramed%
 \at@end@of@kframe}
\makeatother

\definecolor{shadecolor}{rgb}{.97, .97, .97}
\definecolor{messagecolor}{rgb}{0, 0, 0}
\definecolor{warningcolor}{rgb}{1, 0, 1}
\definecolor{errorcolor}{rgb}{1, 0, 0}

\usepackage{alltt}
\usepackage{setspace}
\usepackage{fullpage}
\usepackage{wrapfig}
\usepackage{enumerate}
\usepackage{graphicx}
\usepackage{graphics}
\usepackage{comment}
\usepackage{amsmath,amssymb,amsfonts,amsthm,amsbsy}
\usepackage{stmaryrd} 
\usepackage{verbatim}
\usepackage[natbibapa]{apacite}
\usepackage{soul}
\usepackage{multirow}
\usepackage{makecell}
\usepackage{bm}
\usepackage{hyperref}
\hypersetup{colorlinks=true,
    linkcolor=blue,
    filecolor=blue,
    citecolor=black,
    urlcolor=cyan}
\usepackage{xr}

\newcommand{\maria}{Maria}

\newcommand{\InfPar}[1]{\ensuremath{{#1}_{0}}}
\newcommand{\thetaInf}{\InfPar{\theta}}
\newcommand{\IC}[3][\hat\theta]{\mathrm{IF}_{#1}(#2, #3)}
\newcommand{\bty}{\mathbf{y}_H}
\newcommand{\btY}{\mathbf{Y}_H}

\newcommand{\dt}[3][\theta]{\ensuremath{e_{#1}(#2| {#3})}}
\newcommand{\indicator}[1]{\ensuremath{\mathcal{I}\left[#1 \right]}}
\newcommand{\independent}{\perp}

\newcommand{\PP}{\mathrm{Pr}}
\newcommand{\EE}{\mathbb{E}}
\newcommand{\var}{\mathrm{Var}}

\newcommand{\resid}{e}
\newcommand{\Resid}{E}
\newcommand{\meanDiffT}{d}

\newtheorem{prop}{Proposition}

\newenvironment{ass}[2][Assumption]{\begin{trivlist}
\item[\hskip \labelsep {\bfseries #1}\hskip \labelsep ({\bfseries #2}).]}{\end{trivlist}}

\setcitestyle{notesep={}}

\title{Limitless Regression Discontinuity}

\author{Adam Sales \& Ben B. Hansen\thanks{The authors thank Susan
    Dynarski, Roc\'{i}o Titiunik, Matias Cattaneo, Guido Imbens, Brian Junker,
    Justin McCrary, Walter Mebane, Kerby Shedden, Jeff Smith, the participants in the
    University of Michigan Causal Inference in Education Research
    Seminar and anonymous reviewers for helpful suggestions. They
    also thank Jeffrey Howard and Alexis Santos-Lozada, for sharing
    non-public research replication materials, and John E. Bellquist
    for editing. This
    research was supported by the Institute of Education Sciences,
    U.S. Department of Education (R305B1000012), the U.S. National
    Science Foundation (SES 0753164) and an NICHD center grant (R24 HD041028).
    Any opinions, findings, and conclusions or recommendations expressed in this material are those of the authors. }}
\IfFileExists{upquote.sty}{\usepackage{upquote}}{}
\begin{document} 
\maketitle
\begin{abstract}
Conventionally, regression discontinuity analysis contrasts
a univariate regression's limits as its independent variable, $R$,
approaches a cut-point, $c$, from either side. Alternative methods target the average treatment effect in a small region
around $c$, at the cost of an assumption that treatment assignment,
$\indicator{R<c}$, is ignorable vis a vis potential outcomes.

Instead, the method presented in this paper assumes Residual Ignorability,
ignorability of treatment assignment vis a vis detrended potential
outcomes.  Detrending is effected not with ordinary least squares but with MM-estimation,
following a distinct phase of sample decontamination.  The method's inferences
acknowledge uncertainty in both of these adjustments, despite its
applicability whether $R$ is discrete or continuous; it
is uniquely robust to leading validity threats facing regression
discontinuity designs.

\end{abstract}

\doublespacing

\newpage
\section{Introduction} \label{sec:introduction}

In a regression discontinuity design
\citep[RDD;][]{thistlethwaite1960regression}, treatment is allocated to subjects for
whom a ``running variable'' $R$ exceeds (or falls below) a pre-determined cut-point.
\citet{lee2008randomized} has argued that the regression discontinuity design features ``local
randomization'' of treatment assignment, and is therefore ``a highly
credible and transparent way of estimating program effects''
\citep[][, p. 282]{lee2010regression}.

Take the RDD found in  \citet*[][; hereafter LSO]{lindo2010ability}.
LSO attempt to estimate the effect of
``academic probation,'' an intervention for struggling college students,
 on students' subsequent grade point averages (GPAs).
At one large Canadian university, students with first-year GPAs below
a cutoff were put on probation.
Comparing subsequent GPA ($Y$) for students with
first-year GPA ($R$) just below and above the cutoff should
reveal the effectiveness of the policy at
promoting satisfactory grades.

LSO's data analysis, like that of most RDD studies, used ordinary regression analyses to target an extraordinary parameter.
In Imbens and Lemieux's \citeyearpar{imbens2008regression} telling, for example, the target of estimation is not
the average treatment effect (ATE) in any one region around the cutoff
but rather the ``local'' average treatment effect, or ``LATE'':
the limit of ATEs over concentric ever-shrinking regions, essentially
an ATE over an infinitesimal interval.
Following this ``limit understanding,'' it is common to analyze RDDs
using regression to estimate the functional
relationships of $r$ to $\EE( Y | R=r)$ on either side of the cutoff. The difference between the
two regression functions, as evaluated at the cut-point,
is interpreted as the treatment effect
\citep[e.g.,][]{berk1983capitalizing,angrist1999using}.

However,
the GPAs in the academic probation study are
discrete, measured in 1/100s of a grade point;
hence, limits of functions of GPA do not exist.\footnote{For recent
  methods addressing bias when $R$ is rounded see, e.g.,
  \citet{dong2015regression} and \citet{kolesarRothe17}.
  In those cases, there is a continuous running variable, say $R*$,
  that is unobserved, while observed $R=f(R*)$ for some
  $f(\cdot)$ that may be unknown or non-invertible; then the LATE may be defined in terms of limits of realizations of $R*$. In contrast, in the LSO example $R$ is discrete \emph{by
    definition}.}
Further,
re-analysis of LSO's RDD uncovers evidence of ``social corruption''
\citep{wongWing}---some students appear to have
finely manipulated their GPAs to avoid probation.
This necessitates 
excluding subjects immediately on or around the cut-point---precisely
those students to whom the LATE might most plausibly pertain.
Either circumstance calls
into question the appropriateness of limit-based methods.

\citet{cattaneo2014randomization} base RDD inference on
the model that, in effect, the RDD is a randomized controlled trial
(RCT), at
least in sufficiently small neighborhoods of the cut-point.
Under this assumption, once
attention is confined to such a region, the difference of
$Y$-means between subjects falling above and below the cut-point estimates the ATE within that region.
Despite being natural as a specification of Lee's local randomization concept,
the RCT model involves an independence condition that is rarely plausible in RDDs.
In the LSO example, the data refute this model---unless one
rejects all but the small share of the sample contained in a narrow band
of the cutpoint, sacrificing power and external validity.

To circumvent limitations of the simple RCT model, and of the limit understanding, this paper weds parametric and local
randomization ideas into a novel identifying assumption termed ``residual
ignorability.'' The residual ignorability assumption and corresponding ATE
estimates pertain to all subjects in the data analysis sample; discrete
GPAs do not pose a threat. Manipulation of the running variable
remains a threat, but one that Section~\ref{sec:theMethod}'s combination
of sample pruning and robust M-estimation is uniquely equipped
to address.

The remainder of Section~\ref{sec:introduction} uses a public health
example to introduce residual ignorability and to review distribution-free analysis of RCTs. Limitless RDD analysis
combines these ideas with classical, wholly
parametric methods for RDDs (\S~\ref{sec:robust-analys-covar}),
and RDD specification
tests (\S~\ref{sec:specification}). Section~\ref{sec:theMethod}
adapts residual ignorability to data configurations typical of education studies, and sets out an analysis plan anticipating common challenges
of RDD analysis. Section~\ref{sec:appl-effect-acad} executes the plan
with the LSO study, Section~\ref{sec:simulation}
explores the method's performance in simulations, and
Section~\ref{sec:discussion} takes stock. Replication materials,
including R code, are available as a GitHub repository, \url{https://github.com/adamSales/lrd}.



\subsection{The Death Toll from Hurricane \maria}\label{sec:maria}

Hurricane Maria struck the island of Puerto Rico on September 20,
2017.
In spite of widespread devastation, for nearly a year official
statistics pegged the number of hurricane-induced deaths at just 64.
Estimates from investigative journalists and academic
researchers were higher.
Santos-Lozada and Howard's \citeyearpar{santos2018use} authoritative
analysis considered recorded mortality in
months before and after the hurricane, estimating
Maria to have caused 1,139 deaths in excess of those that would have
occurred otherwise. This section demonstrates the concept of residual ignorability,
if not the scope and particulars of the method detailed in Section~\ref{sec:theMethod},
in a reanalysis of these monthly death counts.

In this example, let $i=1,\dots,12$ denote the months of 2017 and let Puerto
Rico's monthly death counts constitute the outcome, $Y_i$.
The running variable $R_i=i$ is month order and months are
``treated,'' $Z_i=1$,
if and only if $R\ge 9$, Hurricane Maria having occurred in September.
Following \citet{neyman:1923} and \citet{rubin1974estimating}, we may
then take each $i$ to have two potential outcomes: $Y_{Ti}$, a potential
response under the treatment condition (the number of deaths that
would occur were $i$ to fall after Maria); and $Y_{Ci}$, a potential response to
control (the death count were $i$ to fall before Maria).
For each $i$, at most one of $Y_{Ci}$ and $Y_{Ti}$ is
observed, depending on $Z_i$; observed responses $Y$ coincide with
$ZY_{T}+(1-Z)Y_{C}$.
Differences $\tau_i=Y_{Ti}-Y_{Ci}$, $i=9,
\ldots, 12$, represent mortality caused by Maria. We will discuss RDD estimation
of total excess mortality for 2017, $\sum_{i\ge 9} \tau_{i}$, in due
course.
The remainder of this section demonstrates how to test the
hypothesis $\tau_{i}\equiv 0$, all $i$, using a Fisher randomization
test---but without assuming 
the following independence property: 
\begin{ass}{Strong Ignorability; \citealp{rosenbaum1983central}}
\begin{equation}\label{eq:ignore}
Y_{C} \independent Z.
\end{equation}
\end{ass}

\begin{figure}
\centering
\includegraphics[width=0.95\textwidth]{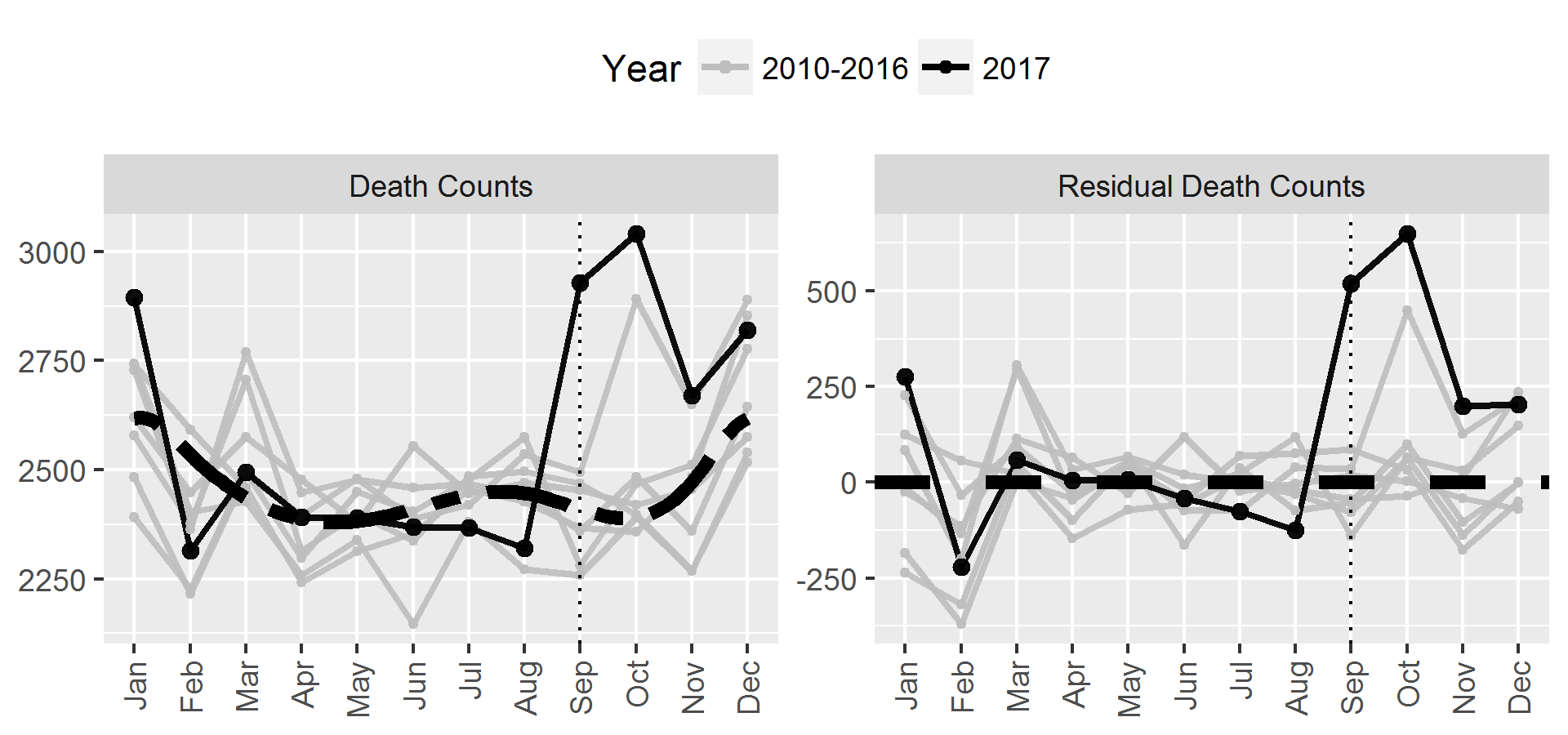}
\caption{Monthly death counts in Puerto Rico from years 2010--2017, before and after
  residualization. The plot on the left shows monthly death counts,
  adjusted for month length. A vertical dotted line denotes September,
  the month \maria\ hit. The fit of the robust model described in the
  text is shown as a dashed black line. The plot on the right shows
  the monthly residuals of the model fit, with a dashed line denoting
  zero.}
\label{fig:maria}
\end{figure}

Although RCTs validate \eqref{eq:ignore} as a
matter of course, the assumption is implausibly strong for the mortality series surrounding
Maria.
For \eqref{eq:ignore} to hold, $Z$ must be independent
of monthly death counts that would have been observed in
the absence of exposure to \maria: the distribution underlying the September through December counts must
be no different than that of the year's first eight months.
Monthly mortality in Puerto Rico for the period 2010--2017, shown in the left panel of
Figure~\ref{fig:maria}, shows there is no precedent for such an equivalence.
Rather, a marked seasonal trend is apparent, with death counts being higher
in the winter months than during the rest of
the year; \eqref{eq:ignore} cannot be sustained.
(For RDD methodology nonetheless founded on \eqref{eq:ignore}, see
\citet{cattaneo2014randomization}
or \citet{matteiMealliObsStud}.)

Dependence between $R$ and $Y_C$, violating
\eqref{eq:ignore}, is common in RDDs; when present, it must be addressed.
Inspection of the 2010--16 mortality series (Figure~\ref{fig:maria}) reveals
a periodic, non-linear relationship
between calendar month and death count, with some years appearing to be
more hazardous than others.
To accommodate these factors, we regressed 2010--2016 monthly death counts
on dummy variables for year and a periodic b-spline for month order,
with knots at February, May, August, and November.
There are several outlying observations, one of which \citet{santos2018use} remove
from the sample prior to analysis.
Rather than identifying and removing outliers informally, we fit the regression model
using a robust redescending M-estimator, which systematically
down-weights and sometimes rejects outliers that would otherwise be
influential \citep{maronna2006robust}.
This model fit is displayed as a dashed black line in Figure
\ref{fig:maria}.

Now let $\hat{Y}_C(R_i)$ be that model's prediction for month $i$ in
2017, and let $\dt[]{Y_i}{R_i} = Y_i-\hat{Y}_C(R_i)$ be the prediction residual,
with potential values $\dt[]{Y_{Ci}}{R_i}=Y_{Ci}-\hat{Y}_C(R_i)$ and
$\dt[]{Y_{Ti}}{R_i}=Y_{Ti}-\hat{Y}_C(R_i)$.
Instead of \eqref{eq:ignore}, we assume only that the model we have fit to
pre-2017 monthly death counts captured and removed seasonal
mortality trends, such that the potential residuals $\{{\Resid}_{Ci}: i\} \equiv \{ \dt[]{Y_{Ci}}{R_i} : i\}$
can be regarded as random, at least as far as $Z$ is concerned:
\begin{equation}\label{eq:ignore2}
\Resid_{C} \independent Z .
\end{equation}
The right panel of Figure \ref{fig:maria} shows residualized death
counts $\Resid = \dt[]{Y}{R}$ as a function of month order $R$.
Seasonal mean trends are no longer in evidence; \eqref{eq:ignore2} is
thus more plausible than the standard ignorability assumption \eqref{eq:ignore}.
Aside from a technical elaboration that will be necessary to apply our method
in the general case (\S~\ref{sec:model-eey-c-r}),
assumption \eqref{eq:ignore2} is residual ignorability, this paper's alternative
to Strong Ignorability as a basis for analysis of RDDs.

\subsection{Using~\eqref{eq:ignore2} to Test the Hypothesis of Strictly No Effect}
\label{sec:using-eqref-test}

In parallel with Fisherian analysis of RCTs \citep{fisher:1935},
which can be regarded as conditioning on the potential outcome random vector $\bm{Y}_C \equiv \{Y_{Ci}: i\}$,
our Maria analysis conditions on the potential residual vector
$\bm{\Resid}_C\equiv\{\Resid_{Ci}\}_{i=1}^{12}$.
(Here and throughout, boldface indicates the concatenation of
$n$ variables or constants.)
The approach applies to the testing of
``strict'' null hypotheses, hypotheses that designate a value for each
$\Resid_{Ti} - \Resid_{Ci}$ with
$Z_{i}=1$, not just $\EE(\Resid_{T} - \Resid_{C} | Z =1)$.
This includes the hypothesis
of strictly no effect, $H_{0}: Y_{T} \equiv Y_{C}$, under which
$\bm{\Resid}_{T} \equiv \bm{\Resid}_{C}=\bm{\Resid}$.

In this analysis $Y$, $R$, and $Z$ data for years 2010--2016 are treated as fixed---formally,
inference will be made after conditioning not only on $\bm{\Resid}_C$ but also on the values of $\{(Y_i, R_i, Z_i) : i \leq 0\}$.
To make use of Fisher's \citeyearpar{fisher:1935} permutation technique,
we likewise condition on the realized sizes $N_{1}$
and $N_{0}$ of the treatment and control group samples,%
\footnote{\textit{Design of Experiments} (\citeyear[, ch.~1]{fisher:1935})
takes sample sizes to be fixed, but Fisher's example of the purple flowers 
(\citealp[e.g.,][]{little:1989,upton1992FET}) demonstrates his view that random $N_{0}$ and $N_{1}$
should be treated as fixed after conditioning upon them.
For quite different assumptions supporting permutation tests in RDDs,
see \citet{canayKamat17RDDpermutations}.
} %
where $N_{j} \equiv \sum_{i=1}^{12} \indicator{Z_{i}=j}$.
Such conditioning is appropriate because the conditioning statistic
$\bm{A}^*=\{\bm{\Resid}_C,N_1,N_0; (Y_{-71}, R_{-71}), \ldots, (Y_{-1}, R_{-1})\}$
is ancillary to, i.e. carries no information about,\footnote{%
$\bm{A}^*$ carries full information about $\bm{\Resid}_{C}$ but none on $\bm{\Resid}_T$.
$\bm{A}^*$ would not be ancillary to targets of the form $\EE(Z| \mathcal{E})$
or $\EE(\Resid_{C}| \mathcal{E})$, some event $\mathcal{E}$.%
} the target of estimation $\EE (\Resid_{T}-\Resid_{C})$.


Under $H_0$, we may exactly enumerate the sampling distribution of any
test statistic $t(\bm{\Resid}_C,\bm{Z})=t(\bm{\Resid},\bm{Z})$ conditional on
$\bm{A}^*$; the permutational p-value is found by comparing a test statistic
to its conditional distribution thus enumerated.
In this analysis $\hat{Y}_C(R_i)$ cannot itself be influenced by \maria,
since it is based on a model fit to pre-\maria\ death counts \citep[cf.][]{rebarPaper}.
Hence, the effect of \maria\ on $\Resid$ (i.e. $\Resid_T-\Resid_C$) is exactly equal to its effect on $Y$.
The null hypothesis $H_0$ states that Hurricane
Maria caused precisely no change to each month's death count, nor to its residual.
Under $H_0$, we condition on $\bm{A}^*$ and calculate the
sampling distribution of the treatment group residual mean $t(\bm{\resid},\bm{Z})=\bar{\resid}_{Z=1}$, by calculating its
value for all $\binom{N_0 + N_1}{N_1} =\binom{12}{4}=495$ possible permutations
of $\bm{Z}$. The null distribution of $t(\bm{\resid}, \bm{Z})$ is simply that of the
mean of a size-4, without-replacement sample from $\{\resid_1, \ldots, \resid_{12}\}$.
It turns out that only 2 permutations of $\bm{Z}$ result in
test statistic values higher than the realized value
$\bar{\resid}_{z=1}=$1,569 (which is unique in
  the distribution).
This implies a two-sided ``mid'' p-value \citep{agresti2005comment} of
$2(2+0.5)/495\approx0.01$
for $H_0$.

This combination of regression and permutation testing applies just as
readily to test the hypothesis $\Resid_T = \Resid_C + c$, for any
constant $c$. In the Maria example, no such hypothesis is sustainable
at level 0.05 unless $c\geq$ 170, corresponding to
680 excess deaths due to the hurricane. Upper
confidence limits and Hodges-Lehmann-type estimates of the effect can
also be obtained in this way. Rather that pursuing this approach
further, we now turn to developing a residual ignorability-based
procedure using M-estimation \citep[][; also called ``generalized
estimating equations'' or ``generalized method of
moments'']{huber1964robust,stefanski2002calculus}, which is better
adapted to data scenarios without the luxury of a separate sample for estimation of
trends in the absence of treatment.

\section{Review of Selected RDD Methods}\label{sec:review}

The method presented in this paper builds on existing methods for RDDs.
This section selectively reviews relevant literature.

Let $Z \in \{0,1\}$ indicate assignment to treatment ($Z=1$) as opposed to control
($Z=0$).
For the remainder of the paper, let $R$ be the \emph{centered} running
variable---the difference between the running variable and the RDD
threshold $c$---so that $Z \equiv \indicator{R< 0}$,
$\indicator{R \leq 0}$, $\indicator{R\geq 0}$ or $\indicator{R > 0}$,
depending on how intervention eligibility relates to the threshold,
where $\indicator{x}=1$ if $x$ is true and $0$ otherwise.
Let $Y$ represent the outcome of interest.
For simplicity assume non-interference, the model that
a subject's response may depend on his but not also on other subjects'
treatment assignments \citep{cox:1958,rubin:1978}. Thus we may take each $i$
to have two potential outcomes, $y_{Ti}$ and $y_{Ci}$, at most one of which is observed;
 observed
responses $Y$ coincide with $ZY_{T}+(1-Z)Y_{C}$.

\subsection{The ANCOVA Model for RDDs}\label{sec:robust-analys-covar}

The classical analysis of covariance (\textsc{ancova}) model for
groups $i=1,\ldots, k$, each including subjects $j=1, \ldots, n_{i}$,
says that
$Y_{ij} = \alpha_{i} + \beta X_{ij} + \epsilon_{ij}$, where $\epsilon_{ij}
\sim \mathrm{Normal}(0, \sigma^{2})$ is independent of the continuous
covariate $X_{ij}$.
In the classical development of RDDs, \textsc{ancova} with $k=2$
groups---treated and untreated---is a leading option among statistical
models
\citep{thistlethwaite1960regression}.
A potential outcomes version of the model is
 $Y_{Ci} = \alpha_{0} + \beta R_{i} + \epsilon_{Ci}$ and
$Y_{Ti} = \alpha_{1}  + \beta R_{i} + \epsilon_{Ti}$, with
 $\epsilon_{Ci} \sim \mathrm{Normal}(0, \sigma^{2})$ and
 $\epsilon_{Ti} \sim \mathrm{Normal}(0, \sigma^{2})$.
In marked contrast to RCTs, it is
not required that $(Y_{T}, Y_{C}) \independent Z$: to the contrary, both
$Y_{C}$ and $Y_{T}$ are presumed to associate with $R$, which in turn
determines $Z$.
Nonetheless, under this model the estimated $Z$ coefficient from the
model
\begin{equation}\label{eq:classicOLS}
Y_i=\alpha+\beta R_i+\tau Z_i+\epsilon_i,
\end{equation}
fit using ordinary least squares (OLS), is unbiased for
$\alpha_{1} - \alpha_{0}$.
Under the \textsc{ancova} model, this estimation target coincides with  $\lim_{r\downarrow 0} \EE(Y | R=r) -
\lim_{r\uparrow 0} \EE(Y | R=r)$ and, simultaneously,
limit-free estimation targets such as $\EE Y_{T}  - \EE Y_{C}$.

The OLS approach estimates $\tau$ as a parameter in regression model
\eqref{eq:classicOLS}. In contrast, the analysis of
\S~\ref{sec:using-eqref-test} took place in two separate steps:
first, adjust outcomes for $R$; then, test hypotheses by contrasting
adjusted outcomes of treated and untreated subjects.
OLS and the \textsc{ancova} model can be also be used for
hypothesis testing, with steps paralleling those of
\S~\ref{sec:using-eqref-test}; this brings an important advantage to
be described in \S~\ref{sec:fuzzy-regr-disc}.
Consider the hypothesis $H: Y_{T} = Y_{C} + \tau$.
Define ${{Y}_H} = {Y} - \tau {Z}$ (so that under $H$, $Y_H=Y_C$)
and $\dt[(a, b)]{\mathbf{y}_{H}}{ \mathbf{r}} = {\mathbf{y}_H} - a -
b\mathbf{r}$.
Finally, test $H$ with statistic
\begin{equation} \label{eq:MeanDiffTestStat}
\meanDiffT(\mathbf{Y}_H , \mathbf{Z}) =
\overline{\dt[(\hat{\alpha},\hat{\beta})]{{{Y}_H}}{ {R}}}_{Z=1} -
\overline{\dt[(\hat{\alpha},\hat{\beta})]{{{Y}_H}}{ {R}}}_{Z=0}
\end{equation}
---where $\hat{\alpha}$ and $\hat{\beta}$ are estimated from an OLS fit of the variant of \eqref{eq:classicOLS}
with dependent variable $\mathbf{y}_{H}$.
A more essential difference between the current section's procedure
and the permutational method of \S~\ref{sec:using-eqref-test} is that the null distribution
of \eqref{eq:MeanDiffTestStat} is not tractable. (In
\S~\ref{sec:using-eqref-test}, test statistics'
permutation distributions were straightforwardly enumerable
because slope and intercept parameters had been estimated
from a separate sample; in \eqref{eq:MeanDiffTestStat}, one cannot
consider alternate realizations of $Z$ without also considering
alternate realizations of $(\hat\alpha,\hat\beta)$.) However, under the parametric \textsc{ancova} model, with
conditioning on $\mathbf{R}$ rather than on $(N_{0}, N_{1},
\mathbf{Y}_{C})$ as in \S~\ref{sec:using-eqref-test},
$\meanDiffT(\mathbf{Y}_H , \mathbf{Z})$ is straightforwardly Normal, with
variance equal to the classical OLS variance of the coefficient on
$Z$.


In general, the set \{$c$:
$H_{c}: Y_{T} = Y_{C} + c$ is not rejected at level $\alpha$\}, which can be seen to be an interval, is a
$100(1-\alpha)\%$ confidence interval for $\tau$ of the Rao score type
\citep{agresti2011scoreintervals}; the $c$ solving
$\meanDiffT(\mathbf{y}_{H}, \mathbf{z}) = 0$, which can be seen to
be unique, is an M-estimate of $\tau$ under both the classical
\textsc{ancova} model and various of its generalizations.
In fact, the estimate for $\tau$ corresponding to these statistical tests is
algebraically equal to the $Z$-coefficient from an OLS estimate of
\eqref{eq:classicOLS}, and the two-sided 95\% confidence interval induced in this
manner is the familiar
$\hat{\tau} \pm 1.96\, \mathrm{SE}(\tau)$.
However, these equivalences do not necessarily extend to estimation
strategies outside of OLS, such as the robust estimators of
\S~\ref{sec:robustFitters} below.

\subsubsection{Addressing the Wald interval's shortcomings for fuzzy RDDs} \label{sec:fuzzy-regr-disc}
RDDs susceptible to non-compliance---where subjects' actual
treatments may differ from $Z$---are called ``fuzzy.''
In these cases, let $D$ indicate whether treatment was actually
received.
This $D$ is an intermediate outcome, so there are
corresponding potential outcomes $D_{C}$ and $D_{T}$, with $D \equiv ZD_{T}
+ (1-Z)D_{C}$.
Subject $i$ is a non-complier if $D_{Ci}=1$ or $D_{Ti}=0$, though we
will assume the monotonicity condition $D_{C}\equiv 0$; there may be
subjects assigned to the treatment who avoid it, but no one gets
treatment without being assigned to it.
We shall also posit the exclusion restriction,
that $Z$ influences $Y$ only by way of its effect on $D$
\citep{bloom1984ans,Angrist:etal:1996,imbens:rose:2005}.
Our focus of estimation is the
 ``treatment-on-treated'' effect (TOTE),
$\EE(Y_{T} - Y_{C}| D_{T}=1)$.

Statistical hypotheses about the TOTE take the form
$H_\tau:Y_T=Y_C+D\tau$.
To test $H_\tau$ under non-compliance, let $Y_H=Y-\tau D$,
designate $t (\mathbf{y}_H,\mathbf{r})$ as test statistic, and compare
its value to a standard Normal distribution.
(The only difference between hypothesis testing for a ``strict'' RDD, one
with full compliance, versus a fuzzy RDD, is in the formulation of
hypothesis $H$, and the construction of $Y_H$---the rest of the
process remains unchanged [\citealp{rosenbaum:1996:onAIR}].)  When
compliance is imperfect, this
iterative method yields confidence intervals with better coverage than Wald-type
confidence intervals---that is, intervals of form $\hat\tau \pm q_{*}
\mathrm{SE}(\hat\tau)$ with $\mathrm{SE}(\hat\tau)$ a single,
hypothesis-independent quantity
\citep[, Sec.~7]{imbens:rose:2005,baiocchiChengSmall2014IVtutorial}.

\subsubsection{Robust Standard Error Estimation}\label{sec:sandwich}
The \textsc{ancova} model for $(Y_{T}, Y_{C})$ is not readily dispensed with, but it
may be relaxed. OLS estimates of $\alpha_1-\alpha_{0}$ and
$\beta$ remain unbiased under non-Normality, provided
the $\epsilon$s have expectation 0 and bounded variances. The
ordinary \textsc{ancova} standard error does not require Normality of
the $( \epsilon_{i}: i )$, either, for use in large samples, although
it does require that they have a common variance.
To test
$\EE\{ \dt[\hat\theta]{{Y_H}}{ R} | Z=1\} = \EE\{ \dt[\hat\theta]{{Y_H}}{
R} | Z=0\}$
under potential heteroskedasticity, one estimates
$\var\left\{\meanDiffT(\mathbf{Y}_H , \mathbf{Z})\right\}$
using a sandwich or Huber-White estimator,
$\mathrm{SE}_{s}^{2} \left\{ \meanDiffT(\mathbf{Y}_H , \mathbf{Z}) \right\}$
\citep{huber1967behavior,mackinnonWhite1985sandwichHC,longErvin2000sandwichHC,
bellmccaffrey2002sandwichSEs,pustejovskyTipton2017sandwichSEs}, 
and refers $\meanDiffT(\mathbf{Y}_H,\mathbf{Z})/\mathrm{SE}_{s}$
to a $t$ or standard Normal reference distribution.
Sandwich standard errors confer robustness to misspecification of
$\var(\dt[\hat\theta]{Y_{H}}{R}\mid  R)$, not of $\EE(Y_{H}| R)$
\citep{freedman2006sch}, the latter being the topic of
the following section.


\subsection{Threats to RDD Validity and some Remedies}\label{sec:specification}

The \textsc{ancova} model for RDDs encodes additional assumptions,
beyond normality and homoskedasticity of regression errors and full
compliance with treatment assignment, which are not so easily
dispensed with.
Methodological RDD literature has responded with specification tests
to detect these threats, or with flexible estimators that
seek to avoid them.

\subsubsection{Covariate Balance Tests}\label{sec:balanceTesting}
Analysis of RCTs and quasiexperiments often hinges on assumptions of
independence of
 $\mathbf{Z}$ from $(\mathbf{X}, \mathbf{Y}_{C}, \mathbf{Y}_{T})$.
 Although neither $\mathbf{Z} \independent \mathbf{Y}_{C}$ nor
 $\mathbf{Z} \independent \mathbf{Y}_{T}$ can be directly tested,
 since potential outcomes are only partly observed, assumptions of form
 $\mathbf{Z} \independent \mathbf{X}$ are falsifiable: researchers can
 conduct placebo tests for effects of $Z$ on $X$.
Of course, treatment cannot affect pre-treatment variables; this is
model-checking (
\citealp[][, \S~5.13]{cox2006pos}
). 

 Writing in the RDD context, \citet{cattaneo2014randomization} test
 for marginal associations of $\mathbf{Z}$ with covariates $\mathbf{X}_{i}$,
 $i=1, \ldots, k$, using the permutational methods that are applied
 in Fisherian analysis of RCTs \citep[also see][]{liMatteiMealli2015BayesianRD}.
Relatedly, \citet{lee2010regression} recommend a
 test for conditional association, given $R$, of $\mathbf{Z}$ and
 $\mathbf{X}$, by fitting models like those discussed in
 \S~\ref{sec:robust-analys-covar} for impact estimation, but with
 covariates rather than outcomes as independent variables.
Viewing the $R$-slopes and intercepts as simultaneously estimated
 nuisance parameters, these are balance tests applied to
 the covariates' residuals, 
rather than to the covariates themselves.

If there are multiple
covariates there will be several such tests. To summarize their
findings with a single p-value, the
regressions themselves may be fused within a
``seemingly unrelated regressions'' model \citep{lee2010regression};
however, to our knowledge, current software implementations do not
support the combination of linear and generalized linear
models, such as when covariates are of mixed type.
Alternatives include hierarchical Bayesian modeling
\citep{liMatteiMealli2015BayesianRD}, or combining
separate tests' p-values using the Bonferroni principle or
other multiple comparison corrections.

\subsubsection{The McCrary Density Test}\label{sec:mccrary}
McCrary's test for manipulation of treatment assignments
\citeyearpar{mccrary2008manipulation} can be understood as a 
placebo test with the density of $R$ as the independent variable.
The test's
purpose is to expose the circumstance of subjects finely manipulating their
$R$ values in order to secure or avoid assignment to treatment. Absent
such a circumstance, if $R$ has a density then it should appear to be
roughly the same just below and above the cutpoint. McCrary's
\citeyearpar{mccrary2008manipulation} test statistic is the difference
in logs of two estimates of $R$'s density at 0, based on observations
with $R<0$ and $R>0$ respectively.
Manipulation is expected to generate a clump just beside the cut
point, on one side of it but not the other, and this in turn engenders
imbalance in terms of distance from the cut-point.

\subsubsection{Reducing the Bandwidth}\label{sec:bandwidth}
In practice, specification test failures inform sample exclusions.
When balance tests fail,
\citet{lee2010regression} would select a bandwidth $b>0$, restrict
analysis to observations with $R\in \mathcal{W} \subseteq [-b, b]$,
and repeat the test on $\{i : r_{i} \in \mathcal{W}\}$.
If that test fails, the process may be repeated with a new bandwidth
$b'<b$, and perhaps repeated again until arriving at suitable bandwidth.
This may seem to call for a further layer of multiplicity correction,
since any number of
bandwidths may have been tested before identifying a suitable
$b$
; but it so happens that this form
of sequential testing implicitly corrects for multiplicity, according to the
sequential intersection union principle
(\citealp[SIUP;][, Proposition~1]{rosenbaum2008testing};
\citealp{hansenSales2015cochran}).
\citet{liMatteiMealli2015BayesianRD} and
\citet{cattaneo2014randomization} also suggest the use of covariate
balance to select a bandwidth.

Restricting analysis to data within a bandwidth may change the
interpretation of the result. The ATE and the
TOTE refer to a discrete population, and reducing the bandwidth likewise
reduces those populations---the new target populations consist of
subjects for whom $|R|\le b$.
(In contrast, the definition of the LATE is unaffected by bandwidth
choice.)


Failures of the density test are addressed by restricting
estimation to observations with $|R|>a$, some $a \geq 0$
\citep[e.g.,][]{barrecaetal2011birthweightRDD,eggers2014validity}, and
repeating the test.
If this test rejects, we repeat the process with a new $a'>a$,
terminating the process when the p-value from the density test
exceeds a pre-set threshold.
By a second application of the SIUP,
the size of this test sequence is equal to the size of each individual
density test.
Taken together, placebo and McCrary tests restrict the sample to
$\mathcal{W} = (-b, b)$ or  $(-b, -a) \cup (a, b)$.

\subsubsection{Non-linear Models for Y as a function of R}\label{sec:nonlinear}
The methods discussed in Sec~\ref{sec:robust-analys-covar} continue to
apply if $\EE (Y_{C}| R) = \alpha + R\beta$ is relaxed to
$\EE (Y_{C}| R) = \alpha + \bm{f}(R)\bm{\beta}$, for
$\bm{f}(\cdot)$ a $1 \times k$ vector valued function, and
$\bm{\beta}$ a $k \times 1$ vector of coefficients.
Unfortunately, if the model is fit by OLS, then such relaxation of assumptions can have the unwelcome
side effect of undercutting the robustness of the analysis. The
reasons have to do with mechanics of regression fitting.

Polynomial specifications
$\EE(Y | R=r) = \sum_{j=0}^{J} r^{j} \beta_{j}$ are common but often
problematic; in combination with ordinary least squares fitting, they
implicitly assign case weights that can vary widely and
counterintuitively \citep{gelman2016high}.
This liability is already
in evidence with $J=1$, the linear specification, where leverage
increases with the square of $r -\bar{r}$.
If analysts select a bandwidth $b$ that is slightly too large, then
the analysis sample will include problematic observations near its outer
boundaries, precisely where leverage is at its highest.
If the analysis sample is contaminated near the cutpoint,
the bad data may not threaten linear specifications, but with $J>1$
they can still bear undue leverage.
In order to identify leverage points that are also influential,
OLS fitting is sometimes combined with specialized diagnostics such as
plots of Cook's \citeyearpar{cook1982residuals} distances.
Section~\ref{sec:robustFitters} will discuss an alternate remedy.


\section{Randomness and Regression in RDDs}\label{sec:theMethod}




The analysis of \S~\ref{sec:maria}
mounted an analogy between the Hurricane Maria RDD and a hypothetical
RCT, but only after a preparatory step of modeling and removing the
outcomes' non-random component.
In \S~\ref{sec:maria}, these two steps used two different
datasets---we regressed $Y$ on $R$ using data from years prior to
2017, when Maria hit, and then used 2017 data to estimate effects,
under the assumption of residual ignorability, \eqref{eq:ignore2}.
This luxury is unavailable in the typical RDD, in which both steps
must use the same data, as in \S~\ref{sec:robust-analys-covar}.
This section will describe a generalization of residual ignorability
\eqref{eq:ignore2} to the typical case, along with robust analysis techniques
incorporating the specification tests reviewed in \S~\ref{sec:specification}.

\subsection{An Analytic Model for RDDs} \label{sec:model-eey-c-r}

This section will formalize residual ignorability for the typical RDD,
which relies on a single dataset including variables $Y$, $R$, and $Z$.
The assumption is that, after a suitable residual transformation,
potential outcomes $Y_C$ are conditionally independent of $Z$.
Hence, causal inference in an RDD may take the perspective
that $Z$ is random due to randomness in $R$.

Suppose the statistician to have selected a \textit{detrending procedure}: a
trend fitter, i.e. a function of
$\{({y}_{i},d_{i},r_{i})\}_{i=1}^{n}$ returning
fitted parameters $\hat{\theta}$ in a sufficiently regular
fashion, along with a
family $\{\dt{\cdot}{\cdot}: \theta\}$ of residual or partial
residual transformations, each mapping data $(\mathbf{y}, \mathbf{r})$ to
residuals
$\{\dt {y_{i}}{{r}_{i}} \}_{i=1}^{n}$.
Appendix~\ref{sec:large-sample-rand}
states the
needed regularity condition, which is ordinarily met by OLS and always
met with our preferred fitters (\S~\ref{sec:robustFitters}).
Then, causal inference in an RDD proceeds from the following assumption:
\begin{ass}{Residual Ignorability}
\sloppy
Given a detrending procedure $(\hat{\theta}, \dt{y}{r})$,
\begin{equation}\label{ycheck}
\dt[\thetaInf]{Y_{C}}{ R }
\independent {Z}| \{R \in \mathcal{W}\}.
\end{equation}
Here $Z = f(R)$, for some deterministic
$f$ (such as $f(r) = \indicator{r < 0}$);  $\thetaInf$ is a constant such that
$\sqrt{n}(\hat\theta - \thetaInf)$ is bounded in probability (and thus
$\hat\theta \stackrel{P}{\rightarrow} \thetaInf$); $\mathcal{W}$
satisfies $\PP( R \in \mathcal{W}) >0$ and $0< \PP(Z=1|R\in\mathcal{W})<1$.
\end{ass}
Residual ignorability states that, though $Y_C$ may not be independent of
$Z$,  it admits a residual transformation bringing about such
independence. With $\dt[\hat\theta]{Y_{C}}{ R}$ a suitable
partial residual, residual ignorability is entailed by the
\textsc{ancova} model (\S~\ref{sec:robust-analys-covar}), or by the combination of any parametric model
for $\EE (Y_{C}| R)$ with a strict null $H$ relative to which the
value of $Y_{C}$ can be reconstructed from the values of $Y$, $D$ and
$Z$ (\S~\ref{sec:using-eqref-test}).
(In either of these cases $\dt[\thetaInf]{Y_{C}}{ R}$
is independent not only of $Z$ but also $R$,
a modest strengthening of~\eqref{ycheck}.)

\sloppy
Assuming residual ignorability, inference about treatment effects is
made conditionally, on
$\mathbf{A}= (\dt[\thetaInf]{\mathbf{Y}_{C}}{ \mathbf{R}}$, $\mathbf{D}_{T},
\{(Y_{Ti}, Y_{Ci}, D_{Ti}, R_{i}) \indicator{{R}_{i} \not\in
  \mathcal{W}}\}_{i=1}^{n})$.
Conditioning on the full data vector when $R \not\in \mathcal{W}$
excludes observations for which \eqref{ycheck} is not assumed.
Conditioning on
$\dt[\thetaInf]{\mathbf{Y}_{C}}{ \mathbf{R}}$
removes little of the randomness 
in $\mathbf{R}$, leaving it available as a basis for inference.
Uncoupled to $Y_{T}$'s, the detrended  $Y_{C}$'s,
$\dt[\thetaInf]{\mathbf{Y}_{C}}{ \mathbf{R}}$,
are in themselves uninformative about $\EE(Y_{T} - Y_{C})$, so
the variables comprising $\mathbf{A}$ are jointly
ancillary, just as $\bm{A}^*$ 
was seen to be
in Section~\ref{sec:using-eqref-test}. As in Fisher-style randomization inference for RCTs, some conditioning variables are
unobserved; but this is not an impediment, at least for large-sample
inferences.

Causal inference based on residual ignorability takes place
in four steps: (1) choosing and validating the analysis sample or
bandwidth, (2) choosing an appropriate fitting procedure (we recommend
robust fitters), (3) treatment effect estimation and inference, and
(4) post-fitting diagnostics.
We will discuss each of these steps in sequence.

\subsection{Pre-Fitting Diagnostics and Bandwidth Choice}
\label{sec:bandwidthChoice}

If subject matter knowledge suggests that the ATE or TOTE would be
most relevant for subjects with $|R|\le b$, then $b$ might form an initial bandwidth choice. But it
is also sensible to subject this choice to specification
testing (\S~\ref{sec:specification}).

Covariate balance or placebo tests for RDDs (\S~\ref{sec:balanceTesting})
assess residual ignorability with a multivariate
``outcome'' $Y^{*}$ combining the actual outcome $Y$ with covariates $X$---\eqref{ycheck} with
$\mathbf{Y}_{C}^{*} = (\mathbf{X}, {Y}_C)$ in place of $Y_{C}$.

Of the placebo testing procedures discussed in
\S~\ref{sec:balanceTesting}, that of \citet{lee2010regression} is best
suited to this conception.
In effect, it begins with preliminary detrending procedures---mechanisms to
decompose  $X$ into components that are systematic or unpredictable,
vis a vis $\mathbf{R}$, just as ${\mathbf{Y}_C}$ will later be decomposed.
Our analysis of the LSO data posits systematic components that are
linear and logistic-linear in $R$, depending on whether $X$ is
a measurement or binary variable. 
The placebo check adds $Z$ to the specification and tests whether its
coefficient is zero. We implement these checks as Wald tests with
heteroskedasticity-robust standard errors, as in
\S~\ref{sec:robust-analys-covar}, using the Bonferroni method to
combine placebo checks across covariates.
To ensure adequate power to detect misspecification,
we test at level $.15$, not $.05$.

We use sequential balance tests to adjust the bandwidth $b$, alongside
McCrary density tests to further
refine the analysis sample $\mathcal{W}$ (\S~\ref{sec:bandwidth}).
These specification tests rely on covariates, $R$, and $Z$, but not on
$Y$; therefore, selection of $\mathcal{W}$ is objectivity-preserving
in the sense of \citet{rubin2007design}.



\subsection{Robust Fitters}\label{sec:robustFitters}
Observations in the analysis sample that do not satisfy residual
ignorability 
 can undercut the validity of an RDD analysis.
Even moderate amounts of such contamination---specifically, contamination of
a $O(n^{-1/2})$-sized
share of the sample that happens to contain influential
observations---can defeat OLS-based estimation strategies, rendering
them inconsistent.
Indeed, even some robust regression methods---those engineered to meet
objectives other than bounding the influence function---may be misled
\citep{stefanski1991note}.
The inclusion of problematic observations in the analysis sample can
be due to misspecification of the model for $\EE (Y_{C}|R)$
(\S~\ref{sec:nonlinear}), manipulation of treatment assignments
(\S~\ref{sec:mccrary}), or other violations of residual ignorability,
coupled with the failure of specification tests to detect these problems.
However, no specification test is powerful enough
to reliably detect moderate contamination; if the probability of a
false alarm is controlled, then power to detect anomalies
affecting only $O(n^{-1/2})$ of the sample can only tend to a number
strictly less than 1.
However $b$ and $\mathcal{W}$ are selected, at least some
contamination may remain in the sample.

Accordingly, consistent estimation of $\thetaInf$ requires robust
M-estimators, in Yohai and Zamar's
\citeyearpar{yohaiZamar1997locallyrobustMestimates} sense, a class
excluding maximum likelihood estimation while including
modern MM-,  SM-, and other
estimators with bounded influence function \citep[see also][, Thm.~3]{he1991localbreakdown}.
 In MM-estimation as in OLS,
coefficients $\bm{\beta}$ of a linear specification solve estimating equations
$\sum_{i} \psi\left\{ ({y}_{i} -
\bm{x}_{i}^T\bm{\beta})/s \right\} \bm{x}_{i}^T =\bm{0}$, where $s>0$ and
$\psi(\cdot)$ is an odd function satisfying $\psi(0)=0$,
$\psi'(0)=1$, and $t\psi(t)\geq 0$; bounded influence fitters replace OLS's $s\equiv 1$ with resistant preliminary
estimates of residual scale and OLS's $\psi(t) = t$ with a continuous $\psi$
that satisfies $\int_{0}^{\infty}\psi(t)dt < \infty$. This limits
the loss incurred by the fitter for failing to adapt itself to a small
portion of aberrant observations;
it is permitted to systematically down-weight them instead.

The analyses and simulations presented below use MM-estimators with bisquare $\psi$ and
``fast S'' initialization \citep{salibian-barreraYohai2006fastS}.
We are not aware of prior work addressing
potential contamination of an RDD sample with the assistance of
bounded influence MM-estimation.
Surprisingly, given their common origins in 
Huber \citeyearpar{huber1964robust}, MM-estimation is not routinely
paired with sandwich estimates of variance, as in
\S~\ref{sec:sandwich} above.
Exceptions include Stata's \texttt{mmregress} and R's \texttt{lmrob},
which optionally provide Huber-White standard errors
\citep{verardiCroux2009robust,rousseuwetal2015robustbase}; our
analyses use the latter.


\subsection{Treatment Effect Estimation and Inference}
\label{sec:test-hypoth-no}

For inference about $\tau$ under the model
$Y_{T} = Y_{C} + \tau D_{T}$, select a specification
$\mu_{\beta}(\cdot)$ for $\EE(Y_{C}| R)$ 
such as the
linear model $\mu_{\beta}(R) =\alpha + R\beta$, a window of
analysis $\mathcal{W}$, and a fitter.

Then, separately for each hypothesis $H: \tau=\tau_0$ under
consideration, calculate
$\mathbf{y}_{H} = \mathbf{y} - \mathbf{d}\tau_{0}$, and
apply the chosen specification and fitter to
$(\mathbf{y}_{H}, \mathbf{r})$.
The combination of the data, the
model fit, and the residual transformation $\dt{\cdot}{\cdot}$ give rise to residuals
$\dt[\hat\theta]{\mathbf{y}_{H}}{\mathbf{r}}$, completing the
detrending procedure. Whether $H$ is rejected or sustained is
determined by the value of the sandwich-based \textsc{ancova} $t$-statistic
in \S~\ref{sec:sandwich}.

In practice it is expedient to use a near-equivalent
test by modifying the detrending
procedure.
When regressing $Y_{H}$ on $R$, include an additive
contribution from $Z$, so that $\mu_{\theta}(R) =\alpha +
R\beta$ is replaced with $\mu_{(\theta,\gamma)}(R) =\alpha +
\beta R + \gamma Z$. With sandwich estimates of
$\text{Cov}\{(\hat{\theta}_{H}, \hat{\gamma}_{H})\}$,
the $t$-ratio comparing $\hat{\gamma}_{H}$ to
$\text{SE}_{s}(\hat{\gamma}_{H})$ induces a generalized score test \citep{boos1992genscoretest}. Implicitly it is a two-sample
$t$-statistic with covariance adjustment for $R$ (with fitting via OLS,
this correspondence would be exact, as noted in Section~\ref{sec:sandwich}; with the robust MM-estimation we
favor, the correspondence is one of large-sample equivalence; see Appendix~\ref{sec:suppl-s-refs}).

As in \S~\ref{sec:robust-analys-covar}, the corresponding
M-estimate of the CACE is the value of $\tau_{0}$ making
$\hat{\gamma}_{H}/\text{SE}_{s}(\hat{\gamma}_{H})$ equal 0; those $\tau_{0}$ for
which $H: \tau = \tau_{0}$ is not rejected at level $\alpha$
constitute a $100(1-\alpha)\%$ confidence interval. Iteration is
facilitated by regressing $\mathbf{y}$ on $\mathbf{r}$ and
$\mathbf{z}$ with offset variable $\mathbf{d}\tau_{0}$; then only the
offset needs to be modified to test $H: \tau = \tau_{1}$,
$\tau_{1}\neq \tau_{0}$.

Strictly speaking, this estimation procedure relies on the assumption of a constant
additive treatment effect, so that
$Y_T=Y_C+D\tau_0$, for some constant $\tau_0$ (or, more generally, an
exact model for treatment effects under which $Y_C$ could be recovered
without error).
This requirement is typical of estimators derived from the inversion
of hypothesis tests \citep[e.g.][]{imbens:rose:2005}.
However, due to its use of sandwich standard errors
$\text{SE}_{s}(\cdot)$, our M-estimator is robust to some
departures from this assumption.
Specifically, under \eqref{ycheck}, $\hat\tau$ is consistent for $\tau_0$ solving
\begin{equation*} 
\EE \big\{\dt[\thetaInf]{Y_T-D_T\tau_0}{ R } \big| R \big\}
\equiv \EE \big\{ \dt[\thetaInf]{Y_{C}}{ R } \big| R \big\},
\end{equation*}
providing such a $\tau_0$ exists.
That is, we assume the existence of a constant $\tau_0$ such that detrended $Y_H=Y-D\tau_0$ is
equal to detrended $Y_C$ on average, if not exactly.
The simulation study in \S~\ref{sec:levelPower} bears out this
robustness property.

\subsection{Post-Fitting Diagnostics} \label{sec:post-fitt-diagn}
Once the M-estimate for the treatment effect has been found, one
inspects the corresponding regression fit for points of high influence.
Robust MM-regression is helpful here. Besides making
influential points easier to see in residual plots, it limits
effects of data contamination, as non-conforming influence points are
down-weighted as a result of the fitting process. This down-weighting
is reflected in ``robustness weights,'' ranging from 1, for non-discounted
observations, down to 0, for the most anomalous observations.
Plotting 
robustness weights against residuals may expose opportunities to
improve the fit of $\mu_{\theta}(R)$, or of the treatment effect model;
plotting them against $R$ may expose contaminated sub-regions
of $\mathcal{W}$ that specification testing failed to remove
\citep{maronna2006robust}.

\section{The Effect of Academic Probation}\label{sec:appl-effect-acad}
\begin{figure}[!ht]
\centering
\includegraphics[width=0.58\textwidth]{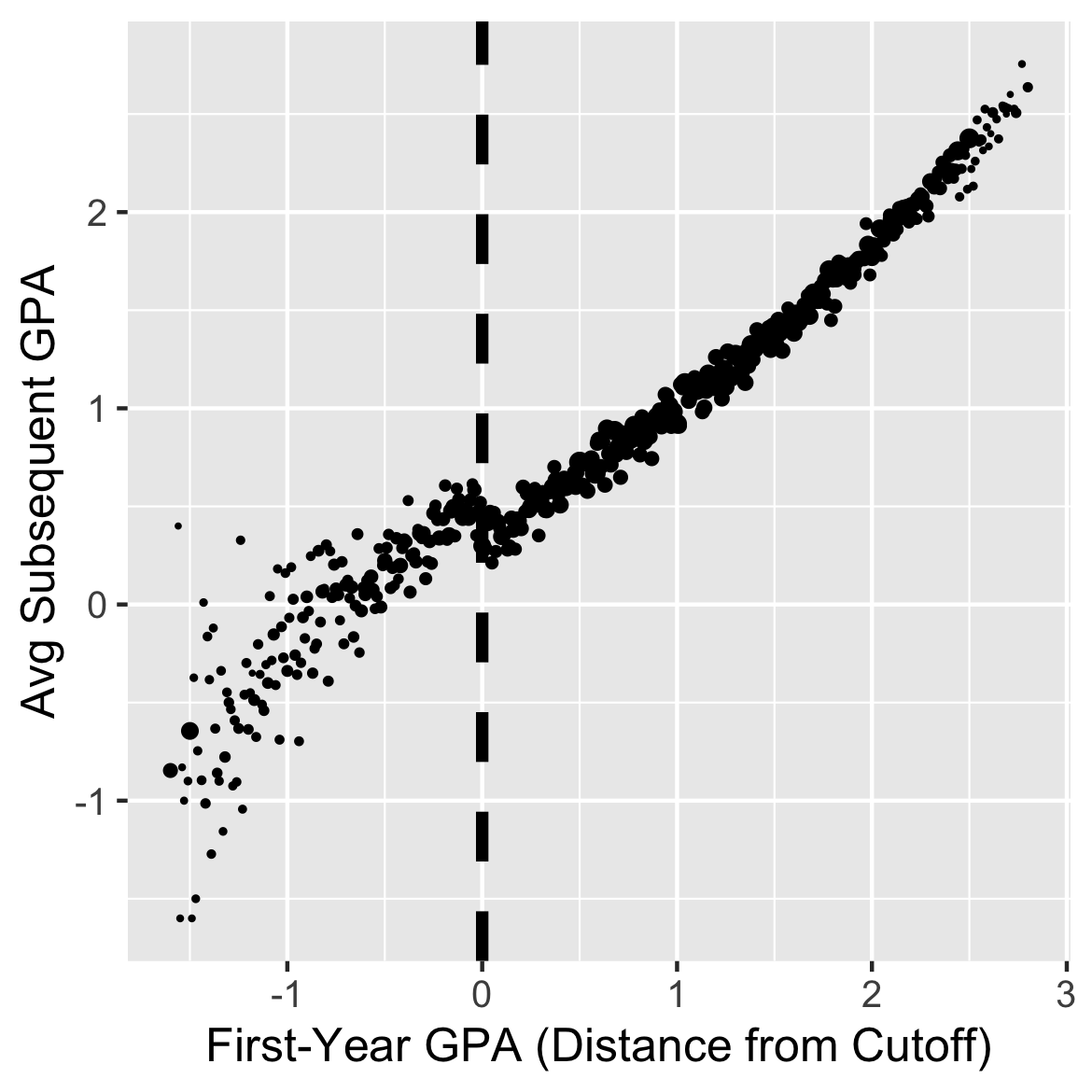}
\caption{
 \texttt{nextGPA} by first-year GPA.
Data are plotted with binning by unique value of first-year GPA, vertical coordinates being bin means of \texttt{nextGPA} and point sizes being proportional to bin size. 
}
\label{LSO}
\end{figure}

Figure~(\ref{LSO}) plots the LSO study's primary outcome,
GPA in the next term a student was enrolled following his
first year (\texttt{nextGPA}), against
first-year GPA.
In all but 50 of 44,362 cases, being on academic probation (AP) coincided with whether
first-year cumulative GPA---the running
variable, $R$---fell below a cutoff.
The university in question had three campuses, two having
cutoffs of 1.5 and the other having a cutoff of 1.6.
To combine data from the three schools, LSO centered each student's
first-year GPA at the appropriate $c$, making $r_i$ the difference of student $i$'s
realized first-year GPA and the cutoff at his or her campus. Figure~\ref{LSO}
follows LSO in this, displaying these $r_{i}$s on its $x$-axis; it also averages \texttt{nextGPA} values over students
with equal first-year GPA, as opposed to plotting students individually.
There is both a discontinuity in \texttt{nextGPA} values as $R$
crosses 0, and a distinctly non-null regression relationship on either side of that
threshold. How large an AP effect may we infer from
these features?  How much of the data bear directly in this inference?

\subsection{Choosing $\mathcal{W}$ and $\mu_{\theta}(\cdot)$} \label{sec:choosing-mathcalw-f}
The region 
$\mathcal{W}_{0.5} =  \pm $ 0.5 
grade points 
includes students whose AP status could change if their
grades in half their classes changed by a full mark (say from D
to C).
Simplicity recommends a linear specification for the outcome
regression on the forcing variable, and the scatter of $Y$ versus $R$
did not suggest otherwise; so we designated
$\mu_{\theta}(R_i)=\alpha+\beta R_i$
and proceeded to specification checks, as
discussed in Section~\ref{sec:specification}.

Following LSO, we conducted placebo tests with high-school
grade percentile rankings, number of credits attempted in first year
of college, first language other than English,
birth outside of North America, age at college entry,
and which of
the university's 3
campuses the student attended.
For the measurement variables, this amounted to fitting
\textsc{ancova} models, whereas binary covariates
were decomposed as logistic-linear in $R$ and $Z$;
in both cases subsequent Wald tests of $Z$'s coefficient used
Huber-White standard errors.
For $\mathcal{W}_{0.5}$  each (Bonferroni-corrected) p-value
exceeds $0.2$; downward adjustment of the bandwidth
is not indicated.


The McCrary density test \citep{mccrary2008manipulation} identifies a
discontinuity in the running variable at the cut-point ($p<0.001$).
AP is a
dubious distinction, and savvy students may try to avoid it. 
Inspection of the distribution of $R$
reveals an unusual number of students whose first-year GPAs were
exactly equal to the AP cutoff, $R=0$.
It would be reasonable to suspect this significant McCrary finding of
being an artifact of the discreteness of first-year GPAs, but
Frandsen's \citeyearpar{frandsenTest} test for manipulation
in a discrete running variable
likewise detected an anomaly at a wide range of tuning parameter values:
$p<0.001$ provided that $0\le k\le0.1$.
The finding is further
corroborated by the fact that the number of students attempting four
or fewer credits was also unusually high in the $R=0$ subgroup,
suggesting that some students dropped courses to dodge AP.
In any event, after removing the $R=0$
subgroup---that is, setting $\mathcal{W}=\{i: |R_i|\in (0,0.5)\} =
\mathcal{W}_{0.5}\setminus \{i: R_{i}=0\}$---the McCrary procedure
narrowly avoids rejecting the hypothesis of no manipulation
($p$ = 0.15).

\subsection{AP Outcome Analysis}
\begin{figure}[ht!]
\centering
\includegraphics[width=.58\textwidth]{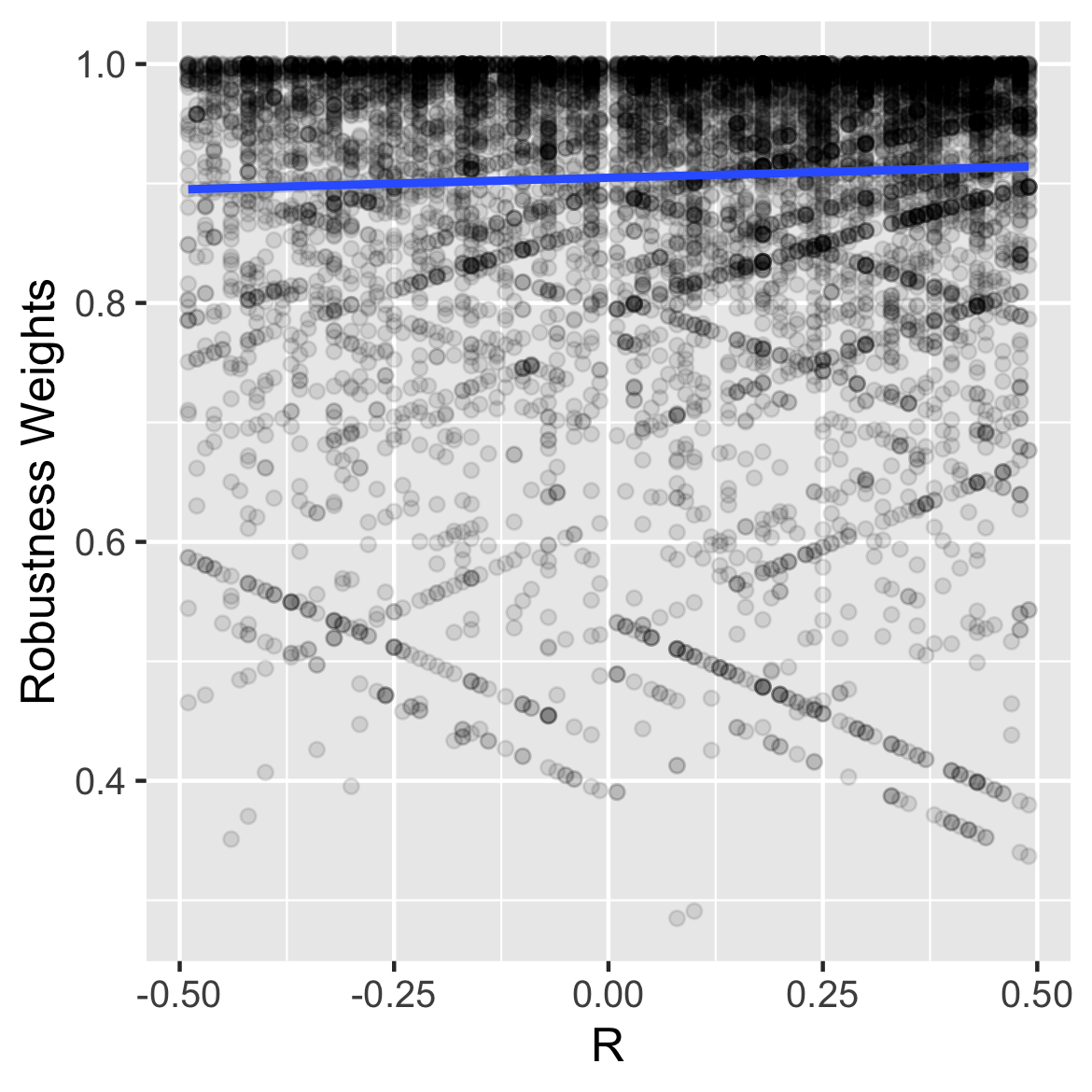}

\caption{Robustness weights from robust MM-estimation of the model
  that $\EE(Y_{C} | R=r)$ is linear in $r$ while $Y_{T} = Y_{C} + \tau_{0} D_{T}$, with $\tau_{0}$ set at $\hat{\tau}=$0.24.}
\label{fig:robweights}
\end{figure}
Table~\ref{tab:results} gives a set of 
estimates for the effect of AP, each obtained
using the robust procedure of \S~\ref{sec:robustFitters}--\ref{sec:test-hypoth-no}.
The first row of Table~\ref{tab:results} gives our main result, with
window of analysis
$\mathcal{W}=\{i: |R_i|\in (0,0.5)\}$ and a linear model of $\EE(Y_C|R)$,
$\mu_\theta(R)=\alpha+\beta R$, estimating the TOTE based on subjects'
received treatments $D$. For the best fitting version
of the model, robustness weights range from .28 to 1.
These weights
show little association with $R$,
although the
lowest weights occur just above the cut-point and near $\mathcal{W}$'s
edges (Figure~\ref{fig:robweights}).

The main analysis estimates an average
treatment effect of 0.24, with 95\% confidence
interval (0.17, 0.31).

Table~\ref{tab:results}'s next three rows relax each of the main model's specifications.
The row labeled ``Adaptive $\mathcal{W}$'' reports the results using
 the wider, adaptively chosen window $\mathcal{W}_{a}=\{i:|R_i|\in
(0,$1.13$)\}$.
The ``Cubic'' row allows for a cubic
relationship between first-year GPA and subsequent GPA, with
$\mu_\theta(R)=\alpha+\beta_1R+\beta_2R^2+\beta_3 R^3$.
This specification performed well in simulations (\S~\ref{sec:simulation}), suggesting that the
warnings in \citet{gelman2016high} against higher-order global
polynomials may not apply to analyses which use robust fitters.
Finally, the ``ITT'' row gives an ``intent to treat'' analysis,
ignoring the difference between students' actual probation and what we
would have expected based on their GPAs.

According to all four analyses, AP gave a modest benefit
over this range.

\begin{table}[ht]
\centering
\begin{tabular}{rcccc}
  \hline
Specification & Estimate & 95\% CI & $\mathcal{W}$ & $n$ \\
  \hline
Main & 0.24 & (0.17, 0.31) & [0.01, 0.50) & 10,014 \\
  Adaptive $\mathcal{W}$ & 0.23 & (0.18, 0.27) & [0.01, 1.13) & 23,874 \\
  Cubic & 0.24 & (0.15, 0.34) & [0.01, 0.50) & 10,014 \\
  ITT & 0.24 & (0.17, 0.31) & [0.01, 0.50) & 10,014 \\
   \hline
\end{tabular}

\caption{AP impact estimates using the method of Section~\ref{sec:theMethod} and
  variants that select $\mathcal{W}$ adaptively, model the outcome
  as a cubic function of the running variable, or estimate
  intent-to-treat effects.}
\label{tab:results}
\end{table}

\subsection{Comparison with Selected Alternatives} \label{sec:simil-diff-with}


For comparison purposes,
we re-analyzed the LSO data using two alternative methods: local
linear regression \citep[e.g.][]{imbens2012optimal}, which targets the difference of limits
of regression functions, and the permutational
method of \citet{cattaneo2014randomization}, which does not
require a limit-based interpretation.
The three sets of results are in Table \ref{tab:alt}.

\begin{table}[ht]
\centering
\begin{tabular}{rcccc}
  \hline
Method & Estimate & 95\% CI & $\mathcal{W}$ & $n$ \\
  \hline
Local Linear & 0.24 & (0.19, 0.28) & [0.00, 1.25) & 26,647 \\
  Limitless & 0.24 & (0.17, 0.31) & [0.01, 0.50) & 10,014 \\
  Local Permutation & 0.10 & (0.04, 0.15) & [0.01, 0.19) & 3,766 \\
   \hline
\end{tabular}

\caption{The effect of Academic Probation from our main analysis compared with permutation and OLS analyses.}
\label{tab:alt}
\end{table}

The local linear approach used the widest window, including
observations with $R<$1.25, and the local
permutation approach used the smallest window, including only
observations with $R<$0.19.
The effect estimates from our method and local linear regression
largely agree, whereas the local permutation approach finds a smaller
effect, with a confidence interval excluding the other two point
estimates.

The data sample for the local linear approach differed from ours in two ways.
First, since the goal of local linear analysis is to estimate regression functions at the
cutoff, it makes little sense to discard observations with $R=c$,
despite counter-indications from the McCrary and Frandsen tests (Section
\ref{sec:choosing-mathcalw-f}).
Second, the \citet{imbens2012optimal} bandwidth is based on non-parametric estimates of the curvature of the
mean function $\EE(Y|R=r)$ rather than on covariates.
We computed this bandwidth using Dimmery's \citeyearpar{rdd}
implementation in \texttt{R}, using the ``rectangular'' kernel option
to facilitate comparisons across methods.
The resulting $\mathcal{W}$ is the widest shown in
Table~\ref{tab:alt}---too wide, from viewpoints either of
Section~\ref{sec:model-eey-c-r}, or of local permutation
analysis. For example, Section~\ref{sec:bandwidthChoice}'s placebo tests reject comparability
of detrended covariate residuals when applied with this $\mathcal{W}$
($p$ = 0.046)
. 

Local linear effect estimation resembles our method,
in that both require analysts to specify and fit models for $Y_C$ and $\tau$.
However, whereas ours calls for robust M-estimation, the local linear method
uses weighted least squares---when the kernel is
rectangular, as in our example, this reduces to OLS within the chosen window.
Confidence intervals are of the Wald type---that is, $\hat{\tau} \pm
z_{\alpha/2} \mathrm{SE}(\hat{\tau})$,
where $z_{\alpha/2}$ is an appropriate normal or $t$-distribution
quantile---rather than inversions of a family of hypothesis tests.
Recent elaborations and extensions include those of \citet{calonico2014robust},
\citet{imbens2017optimized}, and \citet{kolesarRothe17}.

Similar to our approach, the
permutation-based procedure of \citet{cattaneo2014randomization} uses
covariates to select a window of analysis.
However, its covariate balance tests do not adjust for $R$, instead
seeking a $\mathcal{W}$ over which $X \independent Z$ is not
rejected.
In the LSO case, $\mathcal{W}_{b}$ is rejected as long as $b\geq$
0.19. Recall that our $R$-adjusted check found
no fault with bandwidths as large as 1.13.
(In both cases, we tested each $\mathcal{W}$
at level $\alpha=0.15$, addressing multiplicity of covariates using
the Bonferroni method.)

Within the chosen window, the permutational approach estimates effects
under the assumption of
ignorability of treatment assignment, $Z \independent Y_C$.
Failure of this assumption may explain differences between the
permutation-based estimate of the AP effect and estimates from the
other two methods shown in Table~\ref{tab:alt}. A
correlation between \texttt{nextGPA} and $R$---possible even in
regions in which covariate balance cannot be rejected---would
bias a positive effect toward zero.
The Bayesian method of \citet{liMatteiMealli2015BayesianRD}, which
begins from a similar ignorability assumption,
nevertheless models the relationship between $R$ and $Y$ within the
chosen region, to guard against the assumption's failure.
Doing so in the LSO dataset yields a similar point estimate as does the
permutational approach, but with a wider confidence interval that
includes the estimates from our and the local linear approach.

\section{Simulation Studies}\label{sec:simulation}
\subsection{Point and Interval Estimates for Three RDD Methods}\label{sec:levelPower}

Our first simulation study compares the performance---bias and confidence interval
coverage and width---of our ``limitless'' method to local-OLS and
local-permutation methods.
Across all simulation runs, the running variable $R$ was generated as
$\mathrm{Uniform}(-0.5,0.5)$ and control potential
outcomes were generated as $Y_C=0.75R+\epsilon$, where the 0.75 slope was chosen to
approximately match the estimated slope from the LSO study.
Within this framework, we varied three factors: (a) sample size, (b)
the distribution of regression error $\epsilon$, and (c) the
treatment effect.
We considered three sample sizes: $n=50$, 250, and 2,500.
Regression errors were distributed as either Normal or Student's $t$
with 3 d.f.; to mimic the LSO data, we forced the errors to have a
standard deviation of 0.75.
Finally, the treatment effect was either exactly
zero---so $Y_T= Y_C$---or was generated randomly, as $Y_T=
Y_C+\eta$, where $(\sqrt{3}/0.75)\eta\sim t_3$ (so $\eta$
was drawn from the same distribution as was $\epsilon$).
Each simulation scenario was run 5,000
times.

\begin{table}
\footnotesize
\begin{tabular}{ccc|ccc|ccc|ccc}
\hline

&&& \multicolumn{ 3 }{c}{Permutation}&\multicolumn{ 3 }{c}{``Limitless''}&\multicolumn{ 3 }{c}{Local OLS}\\
$n$& Effect& Error & Bias&Cover.&Width&Bias&Cover.&Width&Bias&Cover.&Width \\
\hline 
\hline 
\multirow{3}{*}{ 50 } &0& $\mathcal{N}(0,1)$ &0.37&64&0.91&-0.00&93&1.75&-0.00&93&1.69 \\ 
 &  0 & $t_3$ &0.37&50&0.74&0.01&94&1.41&-0.00&94&1.66 \\ 
 &  $t_3$ & $t_3$ &0.37&65&0.95&0.01&93&1.80&0.01&93&2.04 \\ 
\hline 
\multirow{3}{*}{ 250 } &0& $\mathcal{N}(0,1)$ &0.37&3&0.39&-0.00&95&0.77&-0.00&95&0.75 \\ 
 &  0 & $t_3$ &0.37&0&0.29&-0.00&95&0.57&-0.00&95&0.74 \\ 
 &  $t_3$ & $t_3$ &0.37&3&0.38&-0.00&95&0.73&-0.00&95&0.91 \\ 
\hline 
\multirow{3}{*}{ 2500 } &0& $\mathcal{N}(0,1)$ &0.38&0&0.12&0.00&95&0.24&0.00&95&0.24 \\ 
 &  0 & $t_3$ &0.37&0&0.09&0.00&96&0.17&0.00&95&0.23 \\ 
 &  $t_3$ & $t_3$ &0.37&0&0.11&0.00&94&0.22&0.00&95&0.29 \\ 
\hline
\end{tabular}
  \caption{Empirical bias and 95\% confidence interval coverage (\%) and width for the analyses of 5,000 simulated datasets using either permutation tests, limitless or local OLS methods. The average treatment effect was zero in all conditions; in six conditions the effect was uniquely zero, and in three it was distributed as $t_3$.}
  \label{tab:level}
\end{table}

The results are displayed in Table~\ref{tab:level}.
With a linear data-generating model and a symmetric window, the bias
for the local permutation approach will generally be equal to the
product of the slope and the bandwidth; in our scenario, its bias was
approximately $0.75\times 0.5\approx 0.37$ across simulation runs.
The coverage of permutation confidence intervals decreased with sample size.
The limitless and local OLS methods were approximately unbiased, and
95\% confidence intervals achieved approximately nominal coverage for
$n=250$ or 2,500, and under-covered for $n=50$.
Notably, random treatment effect heterogeneity did not affect bias or
coverage.

Across the board, the local permutation method gave the smallest
confidence intervals; however, this came at the expense of coverage.
Our limitless RD method tended to have equal or slightly narrower
interval widths than the local OLS approach, with greater advantage
when $\epsilon$ was distributed as $t_3$ than when $\epsilon$ was
normally distributed.

\subsection{Polynomial Regression}\label{sec:polynomialSimulation}

        \begin{table}[ht]
\centering
\begin{tabular}{ll|ccccc|ccccc|c } 
  \hline 
&& \multicolumn{5}{c|}{Limitless} &  \multicolumn{5}{c|}{OLS} &Local  \\
 && \multicolumn{5}{c|}{Polynomial Degree}&\multicolumn{5}{c|}{Polynomial Degree}&Linear  \\
 DGM&Measure&1&2&3&4&5&1&2&3&4&5&   \\
\hline
\hline
\multirow{2}{*}{Linear}& bias &0.0&0.0&0.0&0.0&0.0&0.0&-0.0&0.0&0.3&-1.7&0.0\\ 
& RMSE &0.2&0.2&0.3&0.3&0.4&0.3&1.1&4.7&22&106&0.5\\ 
\hline
\hline
\multirow{2}{*}{\makecell[c]{Anti-\\Sym}}& bias &-0.6&-0.6&-0.0&-0.0&0.1&-0.6&1.7&1.8&-9.0&-9.4&-0.0\\ 
& RMSE &0.7&0.7&0.3&0.3&0.4&0.7&2.0&5.0&24&106&0.5\\ 
\hline
\hline
\multirow{2}{*}{Sine}& bias &1.2&1.2&0.2&0.2&0.0&1.2&-2.6&-2.2&1.8&0.2&0.1\\ 
& RMSE &1.2&1.2&0.3&0.3&0.4&1.2&2.9&5.2&21&103&0.5\\ 

 \hline
\end{tabular}
\caption{Results from 5,000 simulations of polynomial specifications for RDD analysis, using limitless, OLS, or local linear regression. Data-generating models (DGM) were as depicted in Figure~\ref{fig:dgms}, with $t_{3}$ errors; sample size for all runs was 500; there was no treatment effect.}
\label{tab:poly}
\end{table}

\begin{figure}
\centering
\includegraphics[width=\maxwidth]{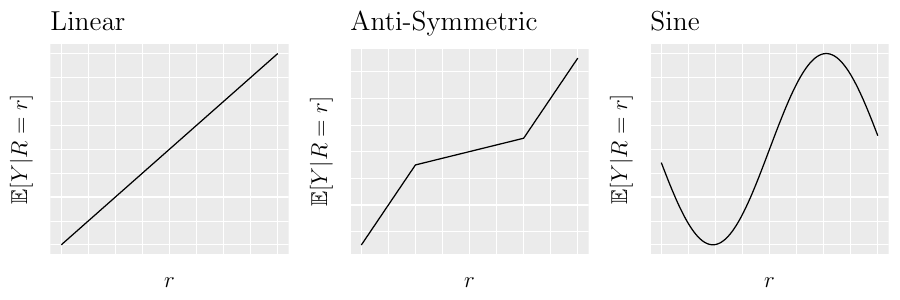} 

\caption{Data-generating models (DGMs) for the polynomial simulation.} 
\label{fig:dgms}
\end{figure}

When $Y$ may not be linear in $R$, flexibility in the
$\mu_{\bm{\theta}}(R)$ function takes on added importance.
We ran an additional simulation to explore the potential of robust
polynomial regression to mitigate influence, as discussed in
\S~\ref{sec:test-hypoth-no} above, while adding flexibility
to the specification of the $Y_{C}$ on $R$ regression.
We compared limitless RD analysis, with $\mu_{\bm{\theta}}$ a polynomial in $R$ with
degree 1, 2, 3, 4, or 5 to analogous estimates from OLS.
In the OLS regressions, we followed the advice of \citet[e.g.][, p. 318]{lee2010regression} and
included interactions between the $R$-polynomial and $Z$.
Finally, we compared these methods to local-linear regression with the
triangular kernel and the bandwidth of \citet{imbens2012optimal}.
The OLS and limitless methods used the entire range of data.
We simulated data sets of size $n=500$ by drawing $R$ and $\epsilon$
from Uniform $(-1,1)$ and  $t_3$ distributions respectively, then
adding $\epsilon$ to one of the three functions of $R$ shown in
Figure~\ref{fig:dgms} to form $Y_{C}$.

Table~\ref{tab:poly} displays the results.
For the linear data-generating model, all estimates were unbiased,
while root mean squared errors (RMSEs) were lowest for the limitless
method.
For the non-linear data-generating models, the limitless estimators
using linear and quadratic specifications had substantial bias, and
bias was much lower for higher-order polynomial specifications.
In contrast, OLS estimators of all polynomial degrees were heavily
biased.
The local linear model does not employ higher-order
polynomials. It fared better than OLS, having similar bias but
higher RMSE than limitless with higher-order polynomials.

OLS and limitless estimation sharply diverge in the quality of their
point estimates, with the OLS estimates' RMSEs exceeding those of
comparable robust M-estimates by factors exceeding 200.
As \citet{gelman2016high} would predict, OLS estimates' RMSEs
increased sharply with each increment of polynomial degree, whatever
the form of the data-generating model.
In marked contrast, under non-linear data-generating models,
higher-degree polynomial terms increased the accuracy of
the robust, limitless method; under linear data-generating models,
including higher-degree terms imposed little penalty.

\section{Discussion} \label{sec:discussion}
Beginning with \citet{thistlethwaite1960regression}, the dominant mode
of RDD analysis has built upon \textsc{ancova} models. A modern
variant instead targets parameters defined
in terms of limits as $r\rightarrow c$, such as the LATE.
However, in RDDs with discrete running variables such as LSO's, neither
$\lim_{r\downarrow 0} \EE (Y |R=r)$ nor $\lim_{r\uparrow 0} \EE (Y | R=r)$
exists, except perhaps as $\EE (Y | R=-.01)$ or $\EE (Y | R=0)$.
A separate embarrassment for  limit-based modeling of RDDs occurs if
a donut-shaped $\mathcal{W}$ is necessary to address potential
manipulation of the running variable, as we found to occur with the
LSO data.

An alternative approach \citep[e.g.][]{cattaneo2014randomization,liMatteiMealli2015BayesianRD} takes the ``local
randomization'' heuristic more literally, analyzing data in a small
region around the cutoff as if it were from a randomized experiment.
However, this approach assumes that potential outcomes are
independent of $R$ in a window around the cutoff.
That assumption is plausible of neither the Hurricane Maria example nor the
LSO case study.  In both settings, it is necessary to acknowledge
and model the $R$--$Y$ relationship in order to set the stage for a
credible claim of independence. The
method of \citet{cattaneo2014randomization}
performed poorly in our simulations
(\S~\ref{sec:simulation}); its discrepant estimate of LSO's AP effect
(Table~\ref{tab:alt}) embodies systematic error.

In contrast, this paper's RDD analysis framework links
\textsc{ancova}- and local randomization heuristics.
Residual ignorability \eqref{ycheck} assumes that the component of
$Y_c$ that depends on $R$ may be modeled and removed, leaving
residuals $\dt[\thetaInf]{\bm{Y}}{\bm{R}}$ that are independent of $Z$.
Like the local randomization approach, it targets the TOTE or ATE
within $\mathcal{W}$, as opposed to a difference of limits, and
accommodates discreteness in $R$ and donut designs.
In the special case of residual ignorability models with
$\dt{\bm{Y}_{C}}{\bm{R}}\equiv\bm{Y}_{C}$ in \eqref{ycheck}, it reduces to the local
randomization method.
Like the limit-based approach, it models and accounts for the correlation between
$R$ and $Y$.
Under certain modeling and fitting choices, it returns the
classical \textsc{ancova} estimate (\S~\ref{sec:robust-analys-covar}).

The method of this paper improves upon each of these approaches by using robust
M-estimation to adjust for $R$.
For analysis of potentially imperfect RDDs, we see this
as a necessity.
For instance, covariate balance tests will necessarily be underpowered
to detect imbalance in a small fraction of the sample, so the proper
bandwidth $b$ will be uncertain.
  Likewise, if the initial sample
includes subjects who manipulated their recorded $R$
values, then the use of donut-shaped $\mathcal{W}$ may
remove some, but not all such subjects. 
Robust
M-estimation retains consistency under scenarios such as these, with moderate amounts of
contamination 
\citep{he1991localbreakdown,yohaiZamar1997locallyrobustMestimates}, whereas OLS does not.

If a large fraction of the dataset violates \eqref{ycheck}, even robust M-estimators
can be misled.
Thus, these methods should be used in addition to, rather than instead of, preliminary specification checks.

In simulated RDDs of moderate size, our estimates were unbiased and
our confidence intervals were typically narrower than those from an
OLS-based approach, while achieving nominal coverage.
Further simulations found robust
M-estimation to be compatible with the use of
global cubic and quartic polynomials to
accommodate nonlinear, imperfectly modeled relationships
between $R$ and $Y_{C}$, in marked contrast to methods using OLS to adjust for
trend.


\bigskip
\singlespacing
{
\bibliographystyle{apacite}
\bibliography{lrd-r1}
}

\doublespacing

\appendix

\section{Large-sample randomization inference for RDDs} \label{sec:large-sample-rand}

In Section~\ref{sec:theMethod} and following, the conditioning is with respect to
\begin{equation} \label{eq:1}
\mathcal{F}^*_{n}=\sigma\left[ \dt[\thetaInf]{\mathbf{Y}_{C}}{
    \mathbf{R}},
\mathbf{D}_{T},  \{(Y_{Ti},
Y_{Ci}, D_{Ti}, R_{i})\indicator{{R}_{i} \not\in
  \mathcal{W}}\}_{i=1}^{n}\right],
\end{equation}
a sigma field bearing
information about $Y_C$s but not $Y_T$s. In contrast to sigma fields
used in Fisherian randomization inference, it does not carry
information about the sizes of the treatment and control groups
(within $\{i : R_{i} \in \mathcal{W}\}$).
In~\eqref{eq:1}, conditioning on full information about subjects $i$
whose $R$ values fell outside $\mathcal{W}$ is a formal reflection of
those subjects' removal from the analytic sample.

As indicated in \S~\ref{sec:maria},
we assume non-interference, the model that
a subject's response may depend on his but not also on other subjects'
treatment assignments \citep{cox:1958,rubin:1978}; for fuzzy RDDs we also
assume the exclusion restriction
and monotonicity (of $(D_{C}, D_{T})$).  In varying degree, the
sections that follow place additional assumptions bounding $Y_{T}$,
$Y_{C}$ or transformations of them.

The Residual Ignorability condition of \S~\ref{sec:model-eey-c-r}
assumes
$\hat\theta$ to be determined in a ``sufficiently regular'' fashion,
noting that bounded influence MM-estimation meets this requirement.  A
weaker regularity condition than boundedness of the influence function
is that the fitter's influence function
$\IC{w}{(\theta, \eta)}$, 
where $(\theta, \eta)$ denotes the full parameter and $w =
({y},d,r)$, must satisfy:
$\EE[\IC{W}{(\theta, \eta)}] =0$ for a unique $\theta = \thetaInf$;
each solution of $\EE[\IC{W}{(\theta, \eta)}] =0$ makes
$\EE[\IC{W}{(\theta, \eta)} \IC{W}{(\theta, \eta)}']$ finite.

\subsection{Distributional approximation for $\hat\theta - \thetaInf$} \label{sec:distr-appr-hatth}
\sloppy
Recall Section~\ref{sec:model-eey-c-r} 
assumes that for the true parameter $(\thetaInf, \InfPar{\eta})$,
$\EE[\IC{W}{(\thetaInf, \InfPar{\eta})}] =0$ and
$\Sigma = \EE[\IC{W}{(\thetaInf, \InfPar{\eta})} \IC{W}{(\thetaInf,
  \InfPar{\eta})}']$ is finite.  As noted by
\citet[\S~3]{stefanski2002calculus}, these entail that
$n^{-1/2} \sum_{i=1}^{n}\IC{W_{i}}{(\thetaInf, \InfPar{\eta})}
\stackrel{d}{\rightarrow} N(0,\Sigma)$ and
$n^{1/2}(\hat\theta - \thetaInf) \stackrel{d}{\rightarrow}
N(0,\Sigma)$, and if fitting is done by MM-estimation also that
sandwich estimates are consistent, for $\Sigma$.  This argument
applies also if $\IC{W_{i}}{(\thetaInf, \InfPar{\eta})}$ is the
influence function of $(\hat\theta, \hat\gamma)$, as opposed only to
$\hat\theta$, where $\hat\gamma$ is a $Z$-coefficient as discussed in
Section~\ref{sec:test-hypoth-no}.

\sloppy
For inference conditioned on $\mathcal{F}_{n}^{*}$ as in \eqref{eq:1},
we require an approximation to the distribution of $\sum_{i=1}^{n} \IC{W_{i}}{(\thetaInf,
    \InfPar{\eta})} - \EE[  \IC{W_{i}}{(\thetaInf,
    \InfPar{\eta})} \vert  \mathcal{F}_{n}^{*} ]$.    For methods
  recommended in this paper
,  it suffices to consider the
  case that $\IC{w}{(\theta,\eta)}$ is bounded.
Write
\begin{equation*}
\mathcal{G}_{n}  = \sigma\left(  \{W_{i}\}_{i=1}^{n};
\{\left(\dt[\thetaInf]{\mathbf{Y}_{C}}{
    \mathbf{R}},
\mathbf{D}_{Ti},  (Y_{Ti},
Y_{Ci}, D_{Ti}, R_{i})\indicator{{R}_{i} \not\in
  \mathcal{W}}\right)\}_{i=n+1}^{\infty}
  \right)
\end{equation*}
so that  $\mathcal{F}_{n}^{*} \subseteq \mathcal{G}_{m}$ for all $n$
and $m$, while $\sum_{i=1}^{n} \IC{W_{i}}{(\thetaInf,
    \InfPar{\eta})} - \EE[  \IC{W_{i}}{(\thetaInf,
    \InfPar{\eta})} \vert  \mathcal{F}_{n}^{*} ]$ is adapted to
  filtration $(\mathcal{G}_{n}: n)$.
Fixing a vector $t$ of the same dimension as $\IC{W_{i}}{(\thetaInf,
  \InfPar{\eta})}$, and writing $M_{n} = \sum_{i=1}^{n} t\IC{W_{i}}{(\thetaInf,
    \InfPar{\eta})} - t  \EE[  \IC{W_{i}}{(\thetaInf,
    \InfPar{\eta})} \vert  \mathcal{F}_{n}^{*} ]$, we see that
  $(M_{n} : n)$ is a martingale. If $\IC{W_{n}}{(\thetaInf,
    \InfPar{\eta})}t$ is $\mathcal{F}_{n}^{*}$-measurable then $M_{n}=0$
  a.s. for all $n$, and is asymptotically $\mathrm{Normal}(0,0^2)$.
Otherwise
  $\EE\{\mathrm{Var}[ t \IC{W_{i}}{(\thetaInf,
    \InfPar{\eta})} \vert \mathcal{F}_{n}^{*}]\} >0$ and $\sum \mathrm{Var}[t   \IC{W_{i}}{(\thetaInf,
    \InfPar{\eta})} \vert \mathcal{F}_{n}^{*}] = O_{P}(n)$.
The Lindeberg condition follows by dominated convergence,
since $\IC{w}{(\theta,\eta)}$ is bounded.
Thus $\{n^{1/2}M_{n}\}$ is asymptotically Normal by L{\'e}vy's
martingale central limit theorem.  Because $t$ was
  arbitrary, it follows that $n^{-1/2}\sum_{i=1}^{n}  t \IC{W_{i}}{(\thetaInf,
    \InfPar{\eta})} - \EE[  t  \IC{W_{i}}{(\thetaInf,
    \InfPar{\eta})}\vert  \mathcal{F}_{n}^{*} ]$ converges in
  distribution to a Normal distribution with mean 0 and
  variance $\var( t \IC{W_{1}}{(\thetaInf,
    \InfPar{\eta})} \mid \mathcal{F}_{1}^{*} )$.  Accordingly
  $n^{1/2}(\hat\theta, \hat\gamma) - (\thetaInf, \InfPar{\gamma})$ is
  asymptotically MVN with covariance
$\mathrm{Cov}[\IC{W_{1}}{(\thetaInf,
    \InfPar{\eta})} \mid \mathcal{F}_{1}^{*} ]$%
.

Sandwich covariance estimates converge to $\Sigma$, not
$\mathrm{Cov}[\IC{W_{1}}{(\thetaInf, \InfPar{\eta})} \mid\mathcal{F}_{1}^{*} ]$;
but since
$t \mathrm{Cov}[\IC{W_{1}}{(\thetaInf, \InfPar{\eta})} \mid \mathcal{F}_{1}^{*} ] t'
\leq t \mathrm{Cov}[\IC{W_{1}}{(\thetaInf, \InfPar{\eta})}] t'$ for
each $t$, under $H$ we have $\hat{\gamma}_{H}/\mathrm{SE}_{s}(\hat{\gamma}_{H})
\stackrel{d}{\rightarrow} \mathrm{Normal}(0, v)$ for some $v\leq 1$ (where ``$\mathrm{SE}_{s}$'' indicates a sandwich estimate).


\subsection{Equivalence of tests from $z$-coefficent and from mean difference of partial residuals} \label{sec:suppl-s-refs}

Let $f_{0}(r)=1$,
$f_1(r) = \indicator{r<0} -\EE \indicator{R<0} = z - p_{z}$, and $f_j(\cdot)$, $j=2,
\ldots, J$ be functions defining  a design matrix for
regression on $Z$ and $R$; denote coefficients of this regression
$\alpha$, $\gamma$, $\beta$, where $\alpha$ and $\gamma$ are scalar
coefficients for $f_{0}(r)$, $f_{1}(r)$ and $\beta$ is a scalar
($J=2$) or column vector ($J>2$) multiplier for row vectors
$\vec{f}(r) = [f_{2}(r)\, \ldots\, f_{J}(r)]$.
We demonstrate that when $\alpha$, $\gamma$, $\beta$ and perhaps other
components of $\theta$ are fitted simultaneously,
in an MM-estimation type regression procedure making their sampling
variability $O_{P}(n^{-1/2})$,
then $\hat\gamma$ differs only by $o_{P}(n^{-1/2})$ from a constant
multiple of $\overline{\dt[\hat\theta]{y_{H}}{r}}_{z=1} - \overline{\dt[\hat\theta]{y_{H}}{r}}_{z=0}$,
the difference in means of partial residuals.

\sloppy
Denote parameters other than $\gamma$, i.e. $\alpha$, $\beta$, a
scale parameter $s$ and perhaps others, collectively by $\eta$, so that
$\theta = (\gamma, \eta)$. Fixing a (strong) hypothesis $H$, write ${w}$  for  $({y}_H,  {r}, z
)$ and $e_{(\gamma, \eta)}(w)$ for $\psi\{ (y_{H} - \alpha -
\gamma f_{1}(r) -
\vec{f}(r) \beta)/s\}$ such that
the estimating functions
are $\Psi_{j}(w, \gamma, \eta) = e_{(\gamma, \eta)}(w) f_j(r)$, $j\leq J$.  The column equation
$\EE \Psi(W, \gamma_{H}, \eta_{H}) =0$, $\Psi(w, \gamma, \eta) = [\Psi_{1}(w, \gamma, \eta), \ldots,
\Psi_{J}(w, \gamma, \eta)]^{t}$,  implicitly defines
$\gamma_{H}$, and defines or contributes to an implicit definition of
$\theta_{H}$.
It's permitted that there be additional estimating equations, not involving
$\gamma$, that contribute to the implicit definition of $\theta$:
in robust regression,  equations defining the preliminary scale
estimate; an equation $\EE Z = p_{z}$ defining $p_{z}$.  Note that
$\Psi_{1}$'s $z$-factor has been centered around $p_{z}$. This facilitates the argument by
ensuring that $\sum_{i}\Psi_{1}(W_{i}, \gamma, \eta) =
np_{z}(1-p_{z}) \{ \overline{e_{(\gamma, \eta)}(W)}_{Z=1} -
\overline{e_{(\gamma,\eta)}(W)}_{Z=0} \}$, but does
not affect the value of $\gamma$ or its estimates $\hat\gamma$, due to
the inclusion via $\Psi_{0}$ of an intercept term. Note that
$\dt[\theta]{{y_H}}{r}  = \dt[(\gamma, \eta)]{{y_H}}{r} = e_{(0, \eta)}(w)$.

\sloppy
Finally, assume that with these estimating equations and $\mathcal{L}(W)$
jointly the estimating function is asymptotically linear,
\begin{multline} \label{eq:3}
n^{-1}\sum_{i=1}^n \Psi(W_i, \hat{\gamma}_{n}, \hat{\eta}_{n}) \\=
n^{-1}\sum_{i=1}^n \Psi(W_i, \gamma_H, \eta_H) +
\dot{\Psi}_H \, [\hat{\gamma}_{n} -\gamma_H, \hat{\eta}_{n} - \eta_H]^{t}
+ o_{P}(n^{-1/2})
\end{multline}
for consistent estimators $(\hat{\gamma}_{n}, \hat{\eta}_{n})$ of $(\gamma_H, \eta_H)$,
where $\dot{\Psi}_H$ is a matrix with rows $\nabla_{\gamma, \eta} \{\EE
\Psi_{j}(W, \gamma, \eta) \}$, as evaluated at ${(\gamma, \eta) = (\gamma_{H},
  \eta_{H})}$.   Assume the fitting procedure generates
consistent estimators of $(\gamma_H, \eta_H)$ and $\dot{\Psi}_H$,
denoted $(\hat\gamma, \hat\eta)$ and
$\hat{\dot{\Psi}}$, with either $\sum_{i=1}^{n}\Psi(W_i, \hat{\gamma},
\hat{\eta}) = 0 $  or at least  $|\sum_{i=1}^{n}\Psi(W_i, \hat{\gamma},
\hat{\eta}) |_{2} = o_{P}(n^{-1/2})$.

For robust regression embodying a residual transformation
$\psi (\cdot )$ that redescends, consistency and \eqref{eq:3} follow
if $\Psi$ is twice differentiable in the parameters, with derivatives
of $\Psi(W, \gamma, \beta, \sigma)$ bounded by a function $K(W)$ with
finite expectation.  This can be assumed of robust regression using
the bisquare or lqq $\psi$ functions, as in R's robustbase package or
Stata's mmregress, if ${Y_H}$ and $(f_j(R) : j)$ have finite
second moments.  If $\Psi$ is the estimating function of an ordinary
or generalized linear model, or robust regression with Huber loss,
suitable conditions for consistency and \eqref{eq:3} are given in
e.g. \citet{he2000parameters}.

If $H$ is true then $\gamma_{H}=0$, and either of $(\hat\gamma,
\hat\eta)$ and $(0, \hat\eta)$  is consistent for $(\gamma_H,
\eta_H) = (0, \eta_{H})$.  Under $H$, then, we can apply \eqref{eq:3}
to either of these estimator sequences, with the consequence that
\begin{align*}
n^{-1}\sum_{i=1}^n \Psi_1(W_i, \hat{\gamma}, \hat{\eta}) - &
                                                             n^{-1}\sum_{i=1}^n \Psi_1(W_i, 0, \hat{\eta})\\
&= \left[\frac{\partial}{\partial\gamma} \EE \Psi_{1}(W,\gamma, \eta)|_{(0, \eta_{H})}\right] \hat{\gamma} + o_{P}(n^{-1/2}) \\
&= {\dot{\Psi}}_{1\gamma} \hat{\gamma} + o_{P}(n^{-1/2}) .\\
\end{align*}
Since $n^{-1}|\sum_{i=1}^n \Psi(W_i, \hat{\gamma}, \hat{\eta})|_{2} =
o_{P}(n^{-1/2})$, this means that up to differences of order
$o_{P}(n^{-1/2})$, $n^{-1}\sum_{i=1}^n \Psi_1(W_i, 0, \hat{\eta})
\approx  \dot{\Psi}_{1 \gamma} \hat\gamma$.  But
$n^{-1}\sum_{i=1}^n \Psi_1(W_i, 0, \hat{\eta}) =
\bar{z}(1-\bar{z})\{\overline{e_{(0,\hat\eta)}(W)}_{Z=1} -
\overline{e_{(0,\hat\eta)}(W)}_{Z=0}\}$, so
\begin{equation*}
n^{1/2}\left| \frac{\bar{z}(1-\bar{z})}{\hat{\dot{\Psi}}_{1\gamma} }
\cdot
\left[
\overline{e_{(0,\hat\eta)}(W)}_{Z=1} -
\overline{e_{(0,\hat\eta)}(W)}_{Z=0} \right]
- \hat\gamma \right|
\stackrel{P}{\rightarrow} 0.
\end{equation*}

\subsection{Differences of residual means involving parameters estimated during detrending}
\label{apnd:requ-forpr-eqref}


Fix $H: Y_T = Y_C + \tau_0 D_T$ and write
\begin{equation}
  \label{eq:yhc}
  {Y}_{HC} = {Y}_{T} - {D}_{T}\tau_{0}\quad
  \text{and}  \quad Y_{H} = ZY_{HC} + (1-Z)Y_{C},
\end{equation}
 so that for each $i$  ${Y}_{HC i}$ is the  $y_{C}$-value that would be
 reconstructed from data $(y_{Ti}, d_{Ti})$ under $H$, if $(y_{Ti},
 d_{Ti})$ rather than $y_{Ci}$ were observed, whereas $\mathbf{y}_{H}$ is the reconstruction of $\mathbf{y}_{C}$
 according to $H$ based on those data that were actually observed.  Thus $\mathbf{Y}_{HC} \equiv \mathbf{Y}_{C}$ if $H$ is true, but not otherwise.
Let $\bar\theta = (\bar\alpha, \bar\beta, \bar{s}, \ldots)$
describe a solution of a system of estimating equations, including a
subsystem $\EE \psi\{ (Y_{H} -
\vec{f}(R) (\alpha, \beta))/s\}\vec{f}(R) =
\mathbf{0}$, some $\vec{f}(\cdot) = [f_{0}(\cdot)\, f_{1}(\cdot)\, \ldots\,
f_{k}(\cdot)]$ and some $\psi(\cdot)$.  (For $n$-vectors $\mathbf{u}$, $\mathbf{v}$, and $\mathbf{x}$,
``$[\mathbf{u}\, \mathbf{v}\, \mathbf{x}]$'' denotes the $n\times 3$ matrix with
columns $\mathbf{u}$, $\mathbf{v}$, and $\mathbf{x}$.)  In the method
described and exemplified in Section~\ref{sec:theMethod} and
subsequent parts of the paper, $\dt[\theta]{y_{H} }{ r} = \psi\{ (y_{H} -
\vec{f}(r) (\alpha, \beta))/s\}$; however, for the arguments presented in
this appendix it is permitted that $\theta$ be estimated using an
unrelated $\psi(\cdot)$ function, prior to and
separately from the residualization $(y,r) \mapsto \dt{y }{ r}$ figuring in tests of $H$.

Given modest regularity conditions on the distribution of
$( Y_{H}, R) = (Z y_{HC} + (1-Z) y_{C}, R)$ (cf. \eqref{eq:yhc})  and the
transformation $\dt{\cdot }{ \cdot}$, one has
\begin{multline}   \label{eq:thetahatdd}
 \{ \overline{\dt[\hat{\theta}]{Y_{H} }{ {R}}}_{Z=1} -
\overline{\dt[\hat{\theta}]{Y_{H} }{ {R}}}_{Z=0} \} -
\{  \overline{\dt[\bar{\theta} ]{ Y_{H} }{ {R}}}_{Z=1} -
\overline{\dt[\bar{\theta}]{ Y_{H} }{ {R}}}_{Z=0}  \}   = \\
\nabla_{\theta} \bigg[ \EE \Big\{
\dt[\theta]{{{Y}_H} }{ {R}}
\Big| Z=1\Big\}   - \EE \Big\{
\dt[\theta]{{{Y}_H} }{ {R}}
\Big| Z=0\Big\} \bigg]_{\theta=\bar\theta}  (\hat\theta - \bar\theta)^{t}  + o_{P}(n^{-1/2}),
\end{multline}
where $\bar\theta = \EE(\hat\theta({\btY}, \mathbf{R}))$;
see Proposition~\ref{prop:thetahatdd} below.   When $\hat\theta$ is
$n^{1/2}$-consistent and asymptotically Normal, this relationship
warrants the use of Huber-White estimates of
$\var\left\{\overline{\dt[\hat{\theta}]{Y_{H} }{ {R}}}_{Z=1} -
\overline{\dt[\hat{\theta}]{Y_{H} }{ {R}}}_{Z=0}\right\}$,
in turn providing a basis for large-sample t-tests.

Approximation~\eqref{eq:thetahatdd} holds for randomized and
quasiexperimental designs alike, but in RCTs the bracketed difference
of expectations vanishes under $H$
\citep{bowers:hans:2008,lin2013agnostic,lin2013agnosticSupp}, and the
product at right of \eqref{eq:thetahatdd} vanishes as well. Up to
$o_{P}(n^{-1})$, the variances
$\var \left\{ \overline{\dt[\hat\theta]{{Y_H} }{ R}}_{Z=1} -
  \overline{\dt[\hat\theta]{{Y_H}}{ R}}_{Z=0} \right\} $
and
$\var \left\{ \overline{\dt[\bar\theta]{{Y_H} }{ R}}_{Z=1} -
  \overline{\dt[\bar\theta]{{Y_H}}{ R}}_{Z=0} \right\}$
are the same, the latter being no greater than the expected value of
the squared standard error of two-sample inference with unequal
variances, as applied to
$\dt[\bar\theta]{{y_H} }{ r}$.  In practice that specific standard error
is unavailable, $\bar\theta$ being unknown; but its ratio with
$\mathrm{SE}_{u}\big\{ \overline{\dt[\theta]{{Y_H} }{ R}}_{Z=1} -
\overline{\dt[\theta]{{Y_H}}{ R}}_{Z=0} \big\} \big|_{\theta =
  \hat{\theta}}$, the unequal-variances two-sample standard error as figured with
substitution of $\hat\theta$ for $\theta$, tends in probability to 1,
by Slutsky's Lemma;
this suffices for the limiting null
distribution of the Studentized $t$-statistic to be standard Normal.
So in an RCT there is no need for any
explicit acknowledgment of sampling variability in $\hat\theta$.

But this argument does not extend to RDDs.  Rather, in an RDD the
expected values compared in \eqref{eq:thetahatdd} coincide only for a
single value, ordinarily 0, of $\theta$'s slope component.  The
differential at right of
\eqref{eq:thetahatdd} is generally nonzero; it must be permitted to
make a contribution. Fortunately any ordinary standard error attaching
to the $z$-coefficient of the regression of ${\mathbf{Y}_{H}}$ on
$\mathbf{1}$, $\mathbf{R}$ and $\mathbf{Z}$ will do so; this includes
Huber-White standard errors.

 Let $\theta \mapsto \dt[\theta]{y}{y} = e_{\theta}(w)$ be
continuously differentiable (for each $w = (y, {r})$).
By the mean value theorem,
\begin{multline} \label{eq:apnd1}
\big\{ [\overline{e_{\hat{\theta} }({w})}_{z=1}
  -\overline{e_{\bar{\theta}}({w})}_{z=1} ] -
      [\overline{e_{\hat{\theta} }({w})}_{z=0}
      -\overline{e_{\bar{\theta}}({w})}_{z=0} ] \big\}\\
  = \nabla_{\theta} \big[\overline{ e_{\theta}({w}) }_{z=1} -
  \overline{ e_{\theta}({w})
  }_{z=0}\big]_{\substack{\theta=\theta^{*} } }
\cdot  (\hat\theta - \bar\theta) ,
\end{multline}
some $\theta^{*}$ on the line segment connecting $\hat\theta$ and
$\bar\theta$.   Of course $\theta^{*} \stackrel{P}{\rightarrow}
\bar\theta$ if $\hat\theta \stackrel{P}{\rightarrow} \bar\theta$.

Let there be a compact neighborhood $\Theta$ of $\bar\theta$ and an
 accompanying envelope function $k_{\Theta}(\cdot)$,  i.e.,
 $|\nabla_{\theta} e_{\theta}(w )| \leq
k_{\Theta}(w)$, all $\theta \in \Theta$ and all $w$, that is integrable, $\EE k_{\Theta}(W) < \infty$.
With this assumption, the uniform strong law
\citep[e.g.,][Ch.16]{ferguson1996largesampletheory} entails that
if $\theta^{*} \stackrel{P}{\rightarrow} \bar\theta$ then
\begin{equation}
  \overline{\big( \nabla_{\theta }e_{\theta}({w})
    |_{\theta=\theta^{*}} \big)}_{Z=z} \stackrel{P}{\rightarrow} \EE
  \big[\nabla_{\theta} e_{\theta}(W) |_{\theta =
    \bar\theta} \big| Z=z\big], \, z=0\, \mathrm{or}\, 1. \label{eq:4}
\end{equation}
Moreover, since there is an integrable envelope function, dominated convergence and the mean value theorem combine to warrant interchanging the expected value and differentiation operations at right of \eqref{eq:4}.

Strengthening the consistency assumption on $\hat\theta$ to root-n consistency, $| \hat\theta - \bar\theta |_{2} = O_{P}(n^{-1/2})$, it now
follows that the difference of the right-hand side of \eqref{eq:apnd1} and
\begin{equation} \label{eq:appendix1}
\nabla_{\theta} \bigg\{\EE \big[
e_{\theta}(W)
\big| Z=1\big]   - \EE \big[
e_{\theta}(W)
\big| Z=0\big] \bigg\}_{\theta=\bar\theta}   \cdot (\hat\theta - \bar\theta)
\end{equation}
is $o_{P}(n^{-1/2})$. This suffices for
\eqref{eq:thetahatdd}.  To summarize:

\begin{prop} \label{prop:thetahatdd} Under (1)--(3) below, $n^{1/2}\{$
  \eqref{eq:appendix1} $-$ RHS of \eqref{eq:apnd1}\}$\stackrel{P}{\rightarrow} 0$ as $n \uparrow
 \infty$, and in consequence \eqref{eq:thetahatdd} holds.
  \begin{enumerate}
  \item for each $(y, r)$, $\nabla_{\theta} \dt[\theta]{y }{
    r}$ exists and is continuous in $\theta$;
  \item for some compact $\Theta$, open $S \subseteq \Theta$ with
    $\bar\theta \in S$, and $k_{\Theta}(\cdot, \cdot)$ with $\EE
    k_{\Theta}({Y_H}, R) < \infty$, $|\nabla_{\theta} \dt[\theta]{y }{ r}| \leq
k_{\Theta}(y, r) $ for all $(y, r)$; and
\item $| \hat\theta - \bar\theta |_{2} = O_{P}(n^{-1/2})$.
  \end{enumerate}
\end{prop}
The argument can be generalized to cover residual transformations that are Lipschitz but not continuously
differentiable in $\theta$, but the generalization is not needed
for ordinary generalized linear model fitters, nor for common
bounded-influence alternatives (e.g., robust regression
with bisquare psi function).








\end{document}